\documentclass{iopart}

\usepackage{graphicx}
\usepackage{array}
\usepackage{iopams} 
\usepackage{bm}
\usepackage[english]{babel}
\usepackage[latin1]{inputenc}

\begin{document}

\title[Extensions of the auxiliary field method]
{Extensions of the auxiliary field method to solve Schr\"{o}dinger equations}

\author{Bernard Silvestre-Brac$^1$, Claude Semay$^2$ and Fabien Buisseret$^2$}

\address{$^1$ LPSC Universit\'{e} Joseph Fourier, Grenoble 1,
CNRS/IN2P3, Institut Polytechnique de Grenoble, 
Avenue des Martyrs 53, F-38026 Grenoble-Cedex, France}
\address{$^2$ Groupe de Physique Nucl\'{e}aire Th\'{e}orique, Universit\'{e}
de Mons-Hainaut, Acad\'{e}mie universitaire Wallonie-Bruxelles, Place du Parc 20,
B-7000 Mons, Belgium}
\eads{\mailto{silvestre@lpsc.in2p3.fr}, \mailto{claude.semay@umh.ac.be}, 
\mailto{fabien.buisseret@umh.ac.be}} 

\begin{abstract}
It has recently been shown that the auxiliary field method is an interesting 
tool to compute approximate analytical solutions of the Schr\"{o}dinger 
equation. This technique can generate the spectrum 
associated with an arbitrary potential $V(r)$ starting from the 
analytically known spectrum of a particular potential $P(r)$. In the 
present work, general important properties of the auxiliary field 
method are proved, such as scaling laws and independence of the results 
on the choice of $P(r)$. The method is extended in order to find 
accurate analytical energy formulae for radial potentials of the form 
$a P(r)+V(r)$, and several explicit examples are studied. Connections existing
between the perturbation theory and the auxiliary field method are also discussed.
\end{abstract}

\submitto{\JPA}
\maketitle

\section{Introduction}

Auxiliary fields, also known as einbein fields, are known for a long time in
quantum field theory. Initially they have been introduced to remove the cumbersome
square roots appearing in relativistic theories. As an example in string field
theory, let us cite the Nambu-Goto Lagrangian which is transformed into the
Polyakov Lagrangian \cite{str}. They are also of common use in many other
fields of physics, such as supersymmetric field theories \cite{susy} and
hadronic physics \cite{af1}.
A particular use of the auxiliary fields is the transformation of a
semi-relativistic kinetic energy term ($\sqrt{\bm{p}^2+m^2}$) appearing in
Salpeter type equations into an apparently non-relativistic one ($\bm{p}^2/(2 \mu)$)
leading to a simpler Schr\"{o}dinger-like equation \cite{sema04}.
Another interesting approach involving auxiliary fields is the transformation
of a problem containing a linear confining term into a new one containing
a harmonic oscillator potential which can lead to analytical expressions \cite{af3}.

Recently, it was realized \cite{sbs} that auxiliary fields can be used as
a tool to obtain analytical approximate expressions for the eigenvalues 
and the eigenstates of a
Schr\"{o}dinger equation. In this work, hereafter labelled SSB, we proposed
a systematic method, called the auxiliary field method (AFM), which gives
approximate expressions for the eigenenergies of
a non-relativistic two-body system interacting through any local and central
potential $V(r)$. For special forms of this potential, an analytical expression
is available. In SSB we studied in particular power-law and logarithmic
potentials and proposed new energy formulae which are much more accurate
than those found in literature up to now.

The search for analytical solutions of the Schr\"{o}dinger equation is very
interesting and the subject of important investigations. In fact, only very
few potentials give rise to an analytical expression for the eigenenergies
valid for any values of the radial quantum number $n$ and orbital quantum
number $l$. The most famous are the harmonic oscillator $V(r)=\frac{1}{2}
\omega r^2$ and the Coulomb potential $V(r)=- \kappa/r$.

Other potentials have analytical solutions but only for $S$-waves (or in
a one-dimensional case). This is in particular the case for the Morse
potential, the Hulthen potential, the Hylleraas potential or the Eckart
potential.
There exist also potentials which are solvable but for particular values
of their parameters. Among them, the Kratzer potential $V(r)=2D(\frac{1}{2}
a^2/r^2 - b/r)$ is famous and analytically solvable in the case $b=a$ \cite{flu}.

But even if an exact analytical expression is not available, it may be very
interesting to have an approximate analytical expression at our disposal.
In addition to a possible benchmark for numerical calculations, it
exhibits the explicit dependence of the energies as a function of the various
parameters and of the quantum numbers. 
Moreover, an analytical expression is always much less time
consuming than the corresponding numerical resolution, and is thus of very
great help in case of a search for a set of parameters for a
potential relying on a chi-square best fit.

Several methods have been invoked to find approximate analytical solutions:
WKB method, semi-classical treatment, variational methods, perturbation
theory, etc. In SSB we showed that the AFM is especially well suited to
pursue this goal and presents very pleasant features. The aim of this
work, which extends the results obtained in SSB, is essentially twofold:
\begin{itemize}
\item to demonstrate a number of very general and interesting properties
of the AFM, in particular the connections with the perturbation theory;
\item to obtain approximate analytical expressions for a wide class of
potentials, some of them being very important and of common use in several
domains of physics.
\end{itemize}

The paper is organized as follows. In the second section, we demonstrate a
number of very general properties concerning the AFM. The third section is
devoted to the application of the general theory to very specific but
important potentials. The fourth section deals with a detailed comparison of
our analytical results and the corresponding numerical values. 
Some conclusions are drawn in the last section.

\section{General properties}
\label{genprop}

\subsection{Principle of the method}
\label{subsec:princmeth}

Our goal is the search for approximate analytical expressions of the eigenvalues
for a two-body non-relativistic Hamiltonian
\begin{equation}
\label{eq:hamilinit}
H=\frac{\bm{p}^2}{2m}+V(r),
\end{equation}
where $m$ is the reduced mass, $r$ the
relative distance between particles, $\bm{p}$ the conjugate momentum 
associated with $\bm{r}$, and $V(r)$ any central potential.
An analytical expression for all the eigenvalues is known explicitly
only for very specific potentials $P(r)$. The basic idea is to rely on such potentials. In
other words, we assume that we are able to obtain an analytical expression
$e(a)$ for the Schr\"{o}dinger equation
\begin{equation}
\label{eq:schrodinit}
\bar{H}(a)  \left| a \right\rangle = \left[ \frac{\bm{p}^2}{2m}+ a P(r)
\right] \left| a \right\rangle = e(a) \left| a \right\rangle,
\end{equation}
in which, at this stage, $a$ is a real parameter.

Our method needs the introduction of
an auxiliary field $\nu$, which is \emph{a priori} an operator. We recall here the
principle of the method. It consists in four steps:
\begin{enumerate}
\item We calculate the function (the prime denotes the derivative with
respect to $r$)
\begin{equation}
\label{eq:funcK}
K(r) = \frac{V'(r)}{P'(r)}
\end{equation}
and denote by $\hat{\nu}$ the value of the auxiliary field which coincides with
this function, 
\begin{equation}
\label{eq:defchaux}
\hat{\nu}=K(r).
\end{equation}
\item We denote by $J=K^{-1}$ the inverse function of $K$. Thus, one has
\begin{equation}
\label{eq:definvK}
r = J(\hat{\nu}).
\end{equation}
Since both $V(r)$ and $P(r)$ do exhibit an analytical form, the same property
holds for $K(r)$. But it is by no means sure that $J(\hat{\nu})$ can be expressed
analytically. The existence of an analytical expression for $J$ is a necessary
condition for our method to obtain analytical expressions as a final result.
\item We build a new Hamiltonian $\tilde{H}$ which depends the auxiliary field
$\nu$ through the following expression
\begin{equation}
\label{eq:Htilde}
\tilde{H}(\nu)= \bar{H}(\nu)+ g(\nu),
\end{equation}
where $\bar H$ is defined by (\ref{eq:schrodinit})
and where the function $g(\nu)$ is given explicitly by
\begin{equation}
\label{eq:defgnu}
g(\nu)=V(J(\nu)) - \nu P(J(\nu)).
\end{equation}
This function $g(\nu)$ makes a bridge between the potential $P(r)$ 
for which an analytical expression
is known and the potential $V(r)$ for which an analytical expression is 
\emph{a priori} not known.
The very important property is that $\hat{\nu}$, coming from (\ref{eq:defchaux}),
cancels the variation of the new Hamiltonian, i.e.
$\delta \tilde{H}(\nu)/ \delta \nu |_{\nu = \hat{\nu}} = 0$. Moreover, one
has the additional crucial identity $\tilde{H}(\hat{\nu}) = H$, as defined
by (\ref{eq:hamilinit}).
\item Considering now $\nu$ no longer as an operator but as a pure number and
taking into account (\ref{eq:schrodinit}), the eigenvalues of 
$\tilde{H}(\nu)$ are
\begin{equation}
\label{eq:enerprop}
E(\nu)=e(\nu)+ g(\nu),
\end{equation}
where $e(\nu)$ are the eigenvalues of $\bar{H}$.
Then, we determine the value $\nu_0$ that minimizes $E(\nu)$: $\partial E(\nu)/
\partial \nu |_{\nu_0}=0$. We propose to consider $E(\nu_0)$ as the approximate
eigenvalues of the Hamiltonian $H$. In SSB, we presented a bound to test the
accuracy of the method. In order to obtain an analytical expression
for these eigenvalues, we must fulfill a second necessary condition: to be able
to determine $\nu_0$ and, then, $E(\nu_0)$ in an analytical way.
\end{enumerate}
This method is completely general and \emph{a priori} valid for any
potential $V(r)$. Nevertheless, in order to get analytical expressions we must
fulfill, as we saw, two conditions: i) first, to be able to invert relation
(\ref{eq:defchaux}) in order to have access to the function $J(\hat{\nu})$ 
defined by (\ref{eq:definvK}) ii) second, to be able to determine $\nu_0$ and to
calculate the corresponding value $E(\nu_0)$ in an analytical way.

An idea for obtaining analytical expressions of the eigenenergies for an arbitrary
potential is the following.
We start with a potential $P(r)=P^{[0]}(r)$ for which the energies of the
corresponding Hamiltonian $H^{[0]}$ are exactly known. We then proceed as above
to find approximate solutions for the eigenenergies of a Hamiltonian $H^{[1]}$
in which the potential is at present $V(r)=P^{[1]}(r)$. In general, a large
class of potentials can be treated in that way. Moreover, by comparison with 
accurate numerical results, we can even refine
the expressions in order to be very close to the exact solution. 

Considering
now these approximate expressions as the exact ones, we apply once more the AFM
with $P(r)=P^{[1]}(r)$ to obtain approximate solutions for the eigenenergies
of a Hamiltonian $H^{[2]}$ in which the potential is at present $V(r)=
P^{[2]}(r)$. Even if analytical solutions for Hamiltonian $H^{[2]}$ were not
attainable directly with $P(r)=P^{[0]}(r)$, it may occur that they indeed are
with $P(r)=P^{[1]}(r)$. Pursuing recursively such a procedure, one can
imagine to get analytical solutions even for complicated potentials.
Presumably, the quality of the analytical expressions deteriorates with the
order of the recursion.

\subsection{Expression of approximate energies}
\label{subsec:expappenerg}
 
Very rare are the potentials $P(r)$ for which an analytical solution is known
for all radial $n$ and orbital $l$ quantum numbers. Among them, the harmonic
oscillator (ho) $P(r)=r^2$ and the Coulomb (C) $P(r)=-1/r$ potentials are
widely used. Taking benefit of this opportunity, the class of 
power-law potentials (pl) has been studied in SSB. Let us consider 
$P^{(\lambda)}(r)= \textrm{sgn}(\lambda) r^\lambda$ with 
$\textrm{sgn}(\lambda)=\lambda/|\lambda|$ and $\lambda \ne 0$. It has been shown 
in SSB that the eigenvalues of the Hamiltonian 
\begin{equation}
\label{eq:plpot}
H^{(\textrm{pl})}_{\lambda}(a) = \frac{\bm{p}^2}{2m}+ a\, \textrm{sgn}(\lambda)
r^\lambda
\end{equation}
can be written under the form
\begin{equation}
\label{eq:eigenerplpot}
e_\lambda^{(\textrm{pl})}(a)=\frac{2+\lambda}{2 \lambda} (a |\lambda|)^{2/(\lambda+2)}
\left ( \frac{N_\lambda^2}{m} \right )^{\lambda/(\lambda+2)},
\end{equation}
where $N_\lambda$ depends on $n$ and $l$ quantum numbers, as well as $\lambda$. 
This formula gives the exact result for the two important situations:
\begin{itemize}
\item the harmonic oscillator potential since in this case $N^{(\textrm{ho})}=
N_{\lambda=2}=2n+l+3/2$;
\item the Coulomb potential for which $N^{(\textrm{C})}=N_{\lambda=-1}=n+l+1$.
\end{itemize}
It has been shown in SSB that, for any physical values of $\lambda$ ($\lambda > -2$), 
a good form for $N_\lambda$ is given by
\begin{equation}
\label{eq:grandNl}
N_\lambda = b(\lambda) n + l + c(\lambda).
\end{equation}
Some functions $b(\lambda)$ and
$c(\lambda)$ were proposed in SSB in order to give an approximation as precise
as $10^{-3}$ for the most interesting (the lowest) values of the quantum numbers
$n$ and $l$. 

Considering the eigenenergies~(\ref{eq:eigenerplpot}) and (\ref{eq:grandNl}) 
as the `exact' ones for
the Hamiltonian~(\ref{eq:plpot}), one can apply the AFM to get the eigenenergies
for the Hamiltonian~(\ref{eq:hamilinit}). Using the recipe given in section
\ref{subsec:princmeth} with $P(r)=P^{(\lambda)}(r)$, we find the following results
\begin{equation}
\label{eq:enu0frompl}
E(\nu_0)=\frac{|\lambda|}{2} \nu_0 J(\nu_0)^\lambda + V(J(\nu_0)),
\end{equation}
the optimal value $\nu_0$ being determined from the equation
\begin{equation}
\label{eq:nu0frompl}
|\lambda|\nu_0 J(\nu_0)^{\lambda+2}=\frac{N_\lambda^2}{m}=Y_\lambda
\end{equation}
and the function $J(\nu)$ coming from the relation
\begin{equation}
\label{eq:Jnufrompl}
|\lambda| \nu J(\nu)^{\lambda - 1}=V'(J(\nu)).
\end{equation}
Let us emphasize that $J(\nu)$ depends only on the potential $V(r)$ and not
on the particular eigenstate we are interested in.
The value $\nu_0$ depends both on the potential (through the $J$ function) and on the
state under consideration (through the $N_\lambda$ quantity).
It is important to stress that the expression resulting from
(\ref{eq:enu0frompl}) suffers from two approximations:
\begin{itemize}
\item The AFM is based on the replacement of an operator $\nu$ by an
optimal value $\nu_0$; this approximation was discussed is detail in SSB and
an estimation of the error was given;
\item Setting for $N_\lambda$ the value (\ref{eq:grandNl}) is also an
approximation whose quality was discussed extensively in SSB.
\end{itemize}
In the case of a harmonic oscillator $\lambda=2$ or a Coulomb potential 
$\lambda=-1$, only the first type of approximation remains.

The application to the two solvable potentials is immediate:
\begin{itemize}
\item For the Coulomb potential ($\lambda=-1$), one deduces
\begin{equation}
\label{eq:enerpropcl}
E^{(\textrm{C})}(\mu_0)=\frac{\mu_0}{2 J(\mu_0)}+ V(J(\mu_0)) =
\frac{\left( N^{(\textrm{C})} \right )^2}{2m J(\mu_0)^2}+ V(J(\mu_0)).
\end{equation}
The function $J(\mu)$ is determined by the condition
\begin{equation}
\label{eq:energunsr}
V'(J(\mu)) J(\mu)^2 = \mu ,
\end{equation}
whereas the value $\mu_0$ is calculated from the transcendental equation
\begin{equation}
\label{detnu0r2}
\mu_0 J(\mu_0) = \frac{\left( N^{(\textrm{C})} \right )^2}{m}.
\end{equation}
\item For the harmonic oscillator ($\lambda=2$), one has similarly
\begin{equation}
\label{eq:enerpropho}
E^{(\textrm{ho})}(\nu_0)=\nu_0 I(\nu_0)^2 + V(I(\nu_0)) =
\frac{\left( N^{(\textrm{ho})} \right )^2}{2m I(\nu_0)^2}+ V(I(\nu_0)).
\end{equation}
The function $I(\nu)$ is determined by the condition
\begin{equation}
\label{eq:energr2}
V'(I(\nu))= 2 \nu I(\nu),
\end{equation}
whereas the value $\nu_0$ is calculated from the transcendental equation
\begin{equation}
\label{detnu0r3}
\nu_0 I(\nu_0)^4 = \frac{\left( N^{(\textrm{ho})} \right )^2}{2m}.
\end{equation}
\end{itemize}
The comparison of the expression for the energies in both cases
(\ref{eq:enerpropcl}) and (\ref{eq:enerpropho}) clearly shows that there
must exist a link between them. This aspect is considered in the next section
and indeed we will prove a very interesting property.

\subsection{Switching from $P^{(\lambda)}(r)$ to $P^{(\eta)}(r)$.}
\label{subsec:switchlmu}

In the preceding section, we derived the approximation $E_{\lambda}$ that can be obtained
for the eigenvalues of the Hamiltonian (\ref{eq:hamilinit}) applying the AFM with
the starting potential $P^{(\lambda)}(r)$. 
Let us assume that, instead of the starting potential $P^{(\lambda)}(r)$, we are
interested by another starting potential $P^{(\eta)}(r)$. The formulae
(\ref{eq:enu0frompl})-(\ref{eq:Jnufrompl}) apply as well, changing $\lambda$ into
$\eta$. In particular, one determines a function $I(\mu)$ depending on the auxiliary
field $\mu$, an optimal value $\mu_0$ depending on a $Y_{\eta}$ quantity, and the
resulting energy $E_{\eta}(\mu_0)$.
In this last approach, let us introduce a new field $\nu$ by the change of variable
\begin{equation}
\label{eq:passlmu}
\nu = \nu(\mu) = \frac{|\eta|}{|\lambda|} \mu I(\mu)^{\eta-\lambda}.
\end{equation}
Using the definition of $I(\mu)$ as a function of the potential $V$, this new
variable can be defined as well as
\begin{equation}
\label{eq:defnewvar}
\nu = \nu(\mu) = \frac{1}{|\lambda|} I(\mu)^{1-\lambda} V'(I(\mu)).
\end{equation}
Using the definition of the $J$ function (\ref{eq:Jnufrompl}), it is easy to show
the relationship
\begin{equation}
\label{eq:linkJI}
I(\mu)=J(\nu(\mu)).
\end{equation}
Defining $\nu_0=\nu(\mu_0)$, a simple calculation shows that
\begin{equation}
\label{eq:nu0frompl2}
|\lambda| \nu_0 J(\nu_0)^{\lambda+2}=\frac{N_\eta^2}{m}=Y_\eta.
\end{equation}
This is exactly the expression~(\ref{eq:nu0frompl})
but in which the quantity $Y_\lambda$ has been replaced by $Y_\eta$.

To achieve the demonstration, let us introduce this value $\nu_0(Y_\eta)$ in the
expression of $E_{\lambda}(\nu_0)$. Using the link between $\nu_0$ and $\mu_0$, it is
easy to show that
\begin{equation}
\label{eq:enu0frompl2}
E_{\lambda}(\nu_0(Y_\eta))=\frac{|\eta|}{2} \mu_0 I(\mu_0)^\eta + V(I(\mu_0))=
E_{\eta}(\nu_0(Y_\eta)).
\end{equation}
In the expression derived from the case $P^{(\lambda)}(r)$, it is sufficient to
change the value $N_{\lambda}$ by $N_\eta$ to obtain the expression derived from
the case $P^{(\eta)}(r)$.
We end up with the very important conclusion that can be stated as a theorem:
\begin{quote}
\emph{If, in the expression $E(N_{\lambda})$ of the approximate energies resulting
from the AFM with $P^{(\lambda)}(r)$, one makes the substitution
$N_{\lambda} \rightarrow N_{\eta}$ (so that $E(N_{\lambda}) \rightarrow
E(N_{\eta})$ with the same functional form for $E$), one obtains the approximate
eigenenergies resulting from the AFM with $P^{(\eta)}(r)$}.
\end{quote}

In a sense, as long as we use a power-law potential $P^{(\lambda)}(r)$ 
as starting potential, there is a universality of the approximate AFM expression of
the eigenvalue, depending only on the potential $V(r)$. The only reminiscence
of the particular chosen potential $P^{(\lambda)}(r)$ 
is the expression of $N_{\lambda}$, as given by (\ref{eq:grandNl}).
This result holds whatever the form chosen for the potential $V(r)$, even if
we are unable to obtain analytical expressions for one case or the other or both.

This property was emphasized in SSB for the particular case of a power-law
potential $V(r)=r^\lambda$ switching from the harmonic oscillator ($\lambda=2$)
to the Coulomb potential ($\lambda=-1$). We proved here that it is in fact
totally general. It is probably related to a well known property in classical
mechanics: one can pass from the motion of a harmonic oscillator to the Kepler
motion by a canonical transformation.

\subsection{Scaling laws}
\label{subsec:scallaw}

Scaling laws represent an important property for non-relativistic Schr\"{o}dinger
equations. They allow to give the expression for the eigenenergies (and wave
functions) of the most general equation in terms of the corresponding eigenenergies
(and wave functions) of a reduced equation which is much simpler to solve.

Let us recall briefly the scaling law for the energy. Let $E(m,G,a)$ be the
eigenvalues of a Schr\"{o}dinger equation corresponding to a system of reduced
mass $m$ subject to a potential of intensity $G$ and characteristic inverse length $a$.
The scaling law gives the relationship between $E(m,G,a)$ and $E(m',G',a')$.
Let us start from the corresponding Schr\"{o}dinger equations
\begin{eqnarray}
\label{eq:scheq1}
\left[ -\frac{1}{2m} \Delta_r + G V(ar) - E(m,G,a) \right] \Psi(\bm{r})=0, \\
\label{eq:scheq2}
\left[ -\frac{1}{2m'} \Delta_r + G' V(a'r) - E(m',G',a') \right]
\Psi'(\bm{r})=0.
\end{eqnarray}
The important point is that it is the same function $V(x)$ which appears in
both equations.
In (\ref{eq:scheq2}), let us make the change of variables 
$\bm{r}=\alpha \bm{x}$ and multiply it by $\chi$. Now, 
we choose the arbitrary parameters $\alpha$ and $\chi$ in order to fulfill the conditions 
$\chi/(m' \alpha^2)=1/m$ and $\alpha a'=a$. In other words, we impose the following values
\begin{equation}
\label{eq:detparamscal}
\alpha = \frac{a}{a'}, \quad \chi = \frac{m'}{m} \left(\frac{a}{a'}
\right )^2.
\end{equation}
With these values, (\ref{eq:scheq2}) can be recast into the form
\begin{equation}
\label{eq:scheq2p}
\fl
\left[ -\frac{1}{2m} \Delta_x + G' \frac{m'}{m} \left(\frac{a}{a'} \right )^2
V(ax) - \frac{m'}{m} \left(\frac{a}{a'} \right )^2 E(m',G',a') \right]
\Psi'\left(\frac{a}{a'} \bm{x}\right)=0.
\end{equation}
Equation~(\ref{eq:scheq1}) can be recovered, 
provided one makes the identification $G = G' (m'/m) (a/a')^2$ and a
similar relation for the energies.

The scaling law is thus expressed in its most general form as
\begin{equation}
\label{eq:scallawgenp}
E(m,G,a) = \frac{m'}{m} \left(\frac{a}{a'} \right )^2 E \left (
m', G'=G \frac{m}{m'} \left(\frac{a'}{a} \right )^2,a' \right ).
\end{equation}
In fact, it is always possible to define the function $V(x)$ so that
$a'=1$. In what follows, and without loss of generality, we will apply
the scaling law for energies under the form
\begin{equation}
\label{eq:scallawgen}
E(m,G,a) = \frac{m'a^2}{m} E \left ( m', G'= \frac{mG}{m' a^2},1 \right ).
\end{equation}
This equality is very powerful since it is valid for the \emph{exact
eigenvalues} of a non-relativistic Schr\"{o}dinger equation based on an
\emph{arbitrary central potential}. It allows to express the energy in terms
of a dimensionless quantity and some dimensioned factors, as we will see below.

The purpose of this section is to prove that the scaling law for energy,
as expressed by (\ref{eq:scallawgen}) still holds for the approximate
expressions derived from a treatment based on auxiliary fields. It was 
observed in SSB but no proof was given. The only
assumption is that the function $P$ is homogeneous, that is we impose the
property $P(ar) = a^p P(r)$ which implies also $P'(ar) = a^{p-1} P'(r)$.
Let us denote here by $M(m,G)$ the eigenvalues of the Hamiltonian
\begin{equation}
\label{eq:hamil0}
\bar H(G) = \frac{\bm{p}^2}{2m} + G P(r).
\end{equation}
In our treatment, the analytical expression for $M(m,G)$ is supposed to be
known. We are searching for the approximate expression of the eigenvalues
of Hamiltonian (\ref{eq:hamilinit}) with $V(r)=G v(ar)$. We call $J$ the
inverse function of $v'/P'$; consequently we have the property 
$v'(J(X)) = X P'(J(X))$.

Let us apply the recipe described in section \ref{subsec:princmeth}.
The function $K(r)=Ga v'(ar)/P'(r)= G a^p v'(ar)/P'(ar)$ must be
identified with the auxiliary field $\hat{\nu}$ and its inversion
provides $r(\hat{\nu})$ with
\begin{eqnarray}
\label{valar}
a r(\hat{\nu}) = J(\hat{\nu}/(G a^p)).
\end{eqnarray}
The auxiliary Hamiltonian to be considered reads
\begin{equation}
\label{eq:auxhamil}
\tilde{H}(\nu)= \frac{\bm{p}^2}{2m} + \nu P(r) + G [v(J(\mu)) - \mu P(J(\mu))]
\end{equation}
with
\begin{equation}
\label{defchmu}
\mu(\nu)=\frac{\nu}{G a^p}.
\end{equation}
The corresponding eigenenergies are thus
\begin{equation}
\label{eq:eigenaux}
E(m,G,a;\nu) = M(m,\nu) + G [v(J(\mu)) - \mu P(J(\mu))].
\end{equation}
The next step now is the determination of the value $\nu_0$ which minimizes
this value of $E$. A simple calculation shows that 
the value of $\mu_0=\mu(\nu_0)$ is explicitly given by
\begin{equation}
\label{eq:detmu0}
a^p M'(m,G a^p \mu_0) = P(J(\mu_0)),
\end{equation}
where $M'(m,\nu)=\partial M(m,\nu)/\partial \nu$.

Lastly, the energy $E(m,G,a)$ we are looking for is just the value $E(m,G,a;
\nu_0)$. The final expression is therefore
\begin{equation}
\label{eq:expenermga}
E(m,G,a) = M(m, G a^p \mu_0) + G [v(J(\mu_0)) - \mu_0 P(J(\mu_0))].
\end{equation}
Since (\ref{eq:scallawgen}) is valid whatever the potential, it
is in particular valid for the energies $M$. One deduces the following relations
\begin{eqnarray}
\label{eq:lienM}
M \left ( m',m G \mu'_0/(m' a^2) \right ) = (m/(m' a^2)) M(m, G a^p \mu'_0)
\end{eqnarray}
and, after differentiation,
\begin{eqnarray}
\label{eq:lienMp}
M' \left ( m',m G \mu'_0/(m' a^2) \right )= a^p M'(m, G a^p \mu'_0).
\end{eqnarray}

Now let us consider the value
\begin{eqnarray}
\label{eq:eigenauxp}
\fl
E(m',G'=(m G)/(m' a^2),1) &=& M(m',m G \mu'_0/(m' a^2)) \\
 &&+(m G)/(m' a^2) [v(J(\mu'_0)) - \mu'_0 P(J(\mu'_0))].
\end{eqnarray}
The value of $\mu'_0$ is obtained from (see (\ref{eq:detmu0}) applied with
$a=1$)
\begin{equation}
\label{eq:detmu0p}
M'(m',m G \mu'_0/(m' a^2)) = P(J(\mu'_0))
\end{equation}
which, because of the property (\ref{eq:lienMp}), can be transformed into
\begin{equation}
\label{eq:detmu0s}
a^p M'(m,G a^p \mu'_0) = P(J(\mu'_0)).
\end{equation}
A comparison between (\ref{eq:detmu0}) and (\ref{eq:detmu0s}) immediately
implies that $\mu'_0=\mu_0$. Thanks to (\ref{eq:lienM}), this last identity
inserted in (\ref{eq:eigenauxp}) proves that
\begin{equation}
\label{eq:scallawgenaux}
E(m,G,a) = \frac{m'a^2}{m} E \left ( m', G'= \frac{mG}{m' a^2},1 \right ).
\end{equation}
But this result is precisely what we expect from the scaling law 
(see (\ref{eq:scallawgen})).

We thus proved that scaling law for energy holds as well for the approximate
expressions derived by the AFM. This result is almost
general in the sense that it is valid whatever the potential under consideration,
even if we are unable to obtain analytical solution. The only restriction is
that the function $P(r)$ is homogeneous. This is of no consequence since the potential 
$P(r)=\textrm{sgn}(\lambda)r^\lambda$ is used in practice.

\subsection{Extension of the method}
\label{subsec:Extens}

Given a potential $V(r)$, the method for obtaining
approximate solutions using auxiliary fields has been presented extensively in
section \ref{subsec:princmeth}. This method can be extended without
difficulty for a Hamiltonian of type 
\begin{equation}
\label{eq:anu0}
H_a=\frac{\bm{p}^2}{2m}+a P(r)+ V(r).
\end{equation}
One introduces an auxiliary
field as before forgetting about the $a P(r)$ contribution. The first 3 steps
of the algorithm remain unchanged. Thus the $\hat{\nu}$ field is the same, as
is the same the function $g(\nu)$. The only difference arises in the expression
(\ref{eq:Htilde}) of $\tilde{H}$ and $\bar{H}$ where $\nu P(r)$ 
has to be replaced by $(a+\nu) P(r)$. As a consequence, 
the corresponding energy (\ref{eq:enerprop}) has to be replaced by
\begin{equation}
\label{eq:newenerg}
E_a(\nu) = e(a+\nu) + g(\nu).
\end{equation}
$E_a(\nu)$ is an eigenvalue of Hamiltonian
\begin{equation}
\label{eq:anu2}
\tilde H_a(\nu) = \bar H(a+\nu)+g(\nu),
\end{equation}
where $\bar H$ is defined by (\ref{eq:schrodinit})
and where $e(a+\nu)$ is an eigenvalue of Hamiltonian $\bar H(a+\nu)$.
An eigenstate of Hamiltonians $\tilde H_a$ and $\bar H(a+\nu)$ is denoted 
$|a+\nu\rangle$, and we have $e(a+\nu) = \langle a+\nu|\bar H(a+\nu) |a+\nu\rangle$.
If $\nu_0$ is the value of $\nu$ which minimizes (\ref{eq:newenerg}), 
then we could expect that 
\begin{equation}
\label{eq:Eanu0}
E_a(\nu_0) = e(a+\nu_0) + g(\nu_0) 
\end{equation}
is a good approximation of $E_a$, an eigenvalue of Hamiltonian~(\ref{eq:anu0}). 
It seems that (\ref{eq:newenerg}) is very similar to (\ref{eq:enerprop}).
Nevertheless, the small difference is important, because, even if the
determination of $\nu_0$ from (\ref{eq:enerprop}) is technically easy, it may
happen that the determination from (\ref{eq:newenerg}) could be much more
involved.

Using the Hellmann-Feynman theorem \cite{feyn} as in SSB, it can be shown that
\begin{equation}
\label{eq:anu3}
\langle a+\nu_0|P(r)|a+\nu_0\rangle = P\left( J(\nu_0) \right). 
\end{equation}
So, $J(\nu_0)$ is a kind of `average point' for the potential $P(r)$. 
If functions $P(r)$ and $V(r)$ are not too different, one could expect that 
\begin{equation}
\label{eq:anu4}
\langle a+\nu_0|V(r)|a+\nu_0\rangle \approx V\left( J(\nu_0) \right),
\end{equation}
and, in particular, that (see (\ref{eq:defchaux}) and (\ref{eq:definvK}))
\begin{equation}
\label{eq:anu4b}
\langle a+\nu_0|\hat \nu|a+\nu_0\rangle \approx \nu_0.
\end{equation}
Within this condition, the optimal value of the constant replacing the auxiliary
field operator 
is close to the mean value of this operator, as mentioned in SSB. 
So, AFM could actually be considered as a `mean field approximation' with
respect to a particular auxiliary field which is introduced to simplify the
calculations.
As it is also shown in SSB, using (\ref{eq:anu3}), one obtains 
\begin{equation}
\label{eq:anu5}
E_a(\nu_0) - E_a \gtrsim V\left(J(\nu_0)\right)-\langle a+\nu_0 | V(r) | a+
\nu_0 \rangle. 
\end{equation}
Except $E_a$, all quantities can be computed analytically in
principle. So an estimation of the accuracy of the approximate eigenvalue
$E_a(\nu_0)$ can be obtained.

It is very instructive to apply (\ref{eq:newenerg}) for the potential 
$P(r)=P^{(\eta)}(r)$. In this case, the function $J(\nu)$ is unchanged
and still given by (\ref{eq:Jnufrompl}) (with the obvious change 
$\lambda \to \eta$), while the new value of $\nu_0$ is given, instead of
(\ref{eq:nu0frompl}), by
\begin{equation}
\label{eq:nu0frompla}
|\eta|(a+\nu_0) J(\nu_0)^{\eta+2}=\frac{N_\eta^2}{m}
\end{equation}
and the corresponding value of the energy by
\begin{eqnarray}
\fl
E(\nu_0)&=&\textrm{sgn}(\eta) \frac{(2+\eta)a+\eta \nu_0}{2}
J(\nu_0)^\eta + V(J(\nu_0)) \nonumber \\
\fl
&=&\textrm{sgn}(\eta) \frac{(2+\eta)a+\eta \nu_0}{2} \left (
\frac{N_\eta^2}{m |\eta|(a+\nu_0)} \right )^{\eta/(2+\eta)}
+V(J(\nu_0)).
\label{eq:enu0frompla}
\end{eqnarray}
In this case, the virial theorem states that
\begin{equation}
\label{eq:virial}
e(a+\nu_0)= \frac{\lambda+2}{2} \langle a+\nu_0 | (a+\nu_0)\textrm{sgn}
(\lambda)r^\lambda | a+\nu_0 \rangle,
\end{equation}
with $e(z)$ given by (\ref{eq:eigenerplpot}). It is then easy to verify
explicitly the equality (\ref{eq:anu3}) with $J(\nu_0)$ given by
(\ref{eq:nu0frompla}). 
These relations also imply that
\begin{equation}
\label{eq:virial2}
\langle a+\nu_0 | r^\lambda | a+\nu_0 \rangle = J(\nu_0)^\lambda.
\end{equation}
But, an outcome of (\ref{eq:anu4}) is 
$\langle a+\nu_0|r|a+\nu_0\rangle \approx J(\nu_0)$ with $J(\nu_0)$ given by 
(\ref{eq:nu0frompla}). So this relation is equivalent to the approximation
$\langle a+\nu_0 | r^\lambda | a+\nu_0 \rangle \approx \langle a+\nu_0 | r |
a+\nu_0 \rangle^\lambda$.

With this in mind, a potential that is worthwhile to be studied is the sum
of two power-law potentials namely
\begin{equation}
\label{eq:sumpowlaw}
V(r)=\textrm{sgn}(\eta) a r^{\eta} + \textrm{sgn}(\lambda) b r^{\lambda}.
\end{equation}
The above considerations show that the $J(\nu)$ function is given by
\begin{equation}
\label{eq:Jnuspl}
J(\nu)=\left ( \frac{|\eta| \nu}{|\lambda| b} \right )^{1/(\lambda-\eta)},
\end{equation}
with the consequence that the optimal value $\nu_0$ is extracted from the
equation
\begin{equation}
\label{eq:nu0spla}
(a+\nu_0) \nu_0^{(\eta+2)/(\lambda-\eta)}=\frac{N_{\eta}^2
(|\lambda| b)^{(\eta+2)/(\lambda-\eta)}}{m |\eta|^{(\lambda+2)/(\lambda-\eta)}}.
\end{equation}
In general, one does not have an analytical expression for the root of such an
equation. This is possible only for very specific values of the powers $\eta$
and $\lambda$.

In this paper, we will study such a problem for the most favourable
cases, where $\eta$ is chosen to give an exact expression for the eigenvalues
(in practice $\eta=2$ and $\eta=-1$) and where $\lambda$ is chosen in order
to have an analytical root.

\subsection{Relation between AFM and perturbation theory}
\label{subsec:relafmpert}

Let us assume that $V(r)=\sigma v(r)$ with $\sigma$ small enough so that
$\sigma v(r) \ll a P(r)$. If $\sigma$ is strictly 0, 
(\ref{eq:anu0}) and (\ref{eq:anu2}) show that $\nu$, and hence $\nu_0$,
vanishes. Switching on the potential gives a non-vanishing but small value
of $\nu_0$. Indeed, $\hat \nu =\sigma v'(r)/P'(r)$, so we can expect that 
$\nu_0 \propto \Or(\sigma)$. Let us remark that $J(0)$ can have a finite value, 
as it can be seen on (\ref{eq:nu0frompla}) for the particular case
$P(r)=P^{(\eta)}(r)$.

>From results of the previous section, we can write
\begin{equation}
\label{eq:pert1}
E_a(\nu) = e(a+\nu) + \sigma v(J(\nu)) - \nu P(J(\nu)).
\end{equation}
Using the definition~(\ref{eq:defchaux}), 
the condition $\partial E_a(\nu)/\partial \nu|_{\nu=\nu_0}=0$ implies that
\begin{equation}
\label{eq:pert2}
\left. \frac{\partial e(a+\nu)}{\partial \nu}\right|_{\nu=\nu_0} =
e'(a+\nu_0)=P(J(\nu_0)).
\end{equation}
It is interesting to compare this relation with (\ref{eq:anu3}). 
Equation (\ref{eq:pert1}) turns then into
\begin{equation}
\label{eq:pert3}
E_a(\nu_0) = e(a+\nu_0) - \nu_0 e'(a+\nu_0) + \sigma v(J(\nu_0)).
\end{equation}
Expanding $e(a+\nu_0)$ and $e'(a+\nu_0)$ in powers of $\sigma$ and keeping only the 
terms $\Or(\sigma)$, we obtain 
\begin{equation}
\label{eq:pert4}
E_a(\nu_0) \approx e(a) + \sigma v(J(\nu_0)).
\end{equation}

For small values of $\sigma$, the contribution of $\sigma v(r)$ 
can also be computed in perturbation. 
With the notations given above, we have then 
\begin{equation}
\label{eq:pert5}
E_{\textrm{pert.}} = \langle a|H_a|a \rangle = e(a) + \sigma \langle a|v(r)|a\rangle,
\end{equation}
where $|a\rangle$ is an eigenstate of the Hamiltonian $\bar H(a)$. Since 
$|a+\nu_0 \rangle$ and $|a\rangle$ differ only by terms $\Or(\sigma)$,
we have (see (\ref{eq:anu3}))
\begin{equation}
\label{eq:pert6}
P\left( J(\nu_0) \right) = \langle a+\nu_0|P(r)|a+\nu_0 \rangle 
= \langle a|P(r)|a\rangle + \Or(\sigma).
\end{equation}
If the condition~(\ref{eq:anu4}) is fulfilled, we can also write
\begin{equation}
\label{eq:pert7}
\langle a|v(r)|a\rangle \approx v\left( J(\nu_0) \right) + \Or(\sigma).
\end{equation}
Using this result in (\ref{eq:pert5}), we see that
$E_{\textrm{pert.}}$ and $E_a(\nu_0)$ differ only by terms $\Or(\sigma^2)$.
Thus, the AFM and the perturbation theory give exactly the
same results at first order provided $V(r)$ do not differ too strongly
from $P(r)$. Let us remark that the perturbation method needs the 
computation of $\langle a|v(r)|a\rangle$. The AFM shows that this calculation
can be replaced by the computation of $v\left( J(\nu_0) \right)$ which could
be simpler in some particular cases.

\section{Application to special potentials}
\label{sec:applic}

In SSB, we exploited the AFM for power-law potentials 
$V(r) = b r^\lambda$. It was shown that an 
analytical expression for the energies exists
for any value of $\lambda$ (we mainly focused our attention 
on values of $\lambda$ comprised between
$-1$ and $+2$) and is given by (\ref{eq:eigenerplpot}). In this section,
we take benefit of the remark of section~\ref{subsec:Extens} to study potentials 
of the form $a P(r)^{(\eta)} \pm b r^\lambda$. As stated above, 
an analytical expression is not necessarily
available. We will examine first in which cases we do have an analytical
expression, and then we will investigate in more details some of them. In the
following, it is assumed that $a > 0$, $b > 0$ and $-2 < \lambda \leq 2$.

\subsection{Solvable potentials}
\label{sec:SolvablePotentials}

Let us examine first the case $P(r)=r^2$. Application of the general method
to this particular case leads to the following equation for the determination
of $\nu_0$ (see (\ref{eq:nu0spla}))
\begin{equation}
\label{eq:detnu0potspec} 
a \pm \nu_0 = X \nu_0^{\frac{4}{2 - \lambda}},
\end{equation}
where $X = X(m,b,N^{(\textrm{ho})})$ is some function of the parameters whose
expression does not matter for our purpose.
This equation is obviously a transcendental equation for which an analytical
solution does not exist automatically. The only cases for which we are sure that
an analytical solution exists is when it can be transformed into a polynomial
of degree less than or equal to 4. In order to investigate this condition, let us
consider $4/(2 - \lambda)=p/q$ as a rational number ($p$ and $q$ are relatively 
prime). Calling $x=\nu_0^{1/q}$, (\ref{eq:detnu0potspec}) is recast as
\begin{equation}
\label{eq:detnu0potspec1} 
a \pm x^q = X x^p.
\end{equation}
All the solvable potentials should verify the conditions $1 \leq p \leq 4$,
$1 \leq q \leq 4$. An exhaustive research of all the solvable potentials
leads to the following values of the power $\lambda$:
\begin{equation}
\label{vallambr2}
\lambda = -2,\: -1,\: -\frac{2}{3},\: \frac{2}{3}, \: 1.
\end{equation}
We will study in detail the cases:
\begin{itemize}
\item $\lambda = -2$ because it corresponds to a centrifugal term with a
real parameter instead of the usual $l(l+1)$ term;
\item $\lambda = - 1$ because it corresponds the a simplified potential
for hadronic systems with a short range Coulomb potential and a quadratic
confinement;
\item $\lambda = 1$ because this anharmonic potential is sometimes used
in molecular physics.
\end{itemize}

Now we investigate the case $P(r)= -1/r$. The equation corresponding to the
determination of $\nu_0$ then reads for $-1 < \lambda \leq 2$ (see
(\ref{eq:nu0spla}))
\begin{equation}
\label{eq:detnu0potspec2} 
(a \pm \nu_0) \nu_0^{\frac{1}{1+\lambda}}=Z,
\end{equation}
where again $Z=Z(m,b,N^{(\textrm{C})})$ is some unimportant function.
Introducing again the integers $p$ and $q$ such that $1/(1+\lambda)=p/q$ and
$x=\nu_0^{1/q}$, (\ref{eq:detnu0potspec2}) becomes
\begin{equation}
\label{eq:detnu0potspec3} 
(a \pm x^q)x^p = Z.
\end{equation}
All the possibilities to choose $p$ and $q$ such that $p+q \leq 4$ are suitable.
The list of solvable potentials is given below:
\begin{equation}
\label{vallamb1sr}
\lambda = -\frac{2}{3},\:-\frac{1}{2}, \: 1, \: 2.
\end{equation}
Among them we will study:
\begin{itemize}
\item $\lambda = 2$ because it corresponds to a potential $-a/r + b r^2$ which
is already studied with $P(r)=r^2$. This is the only potential that can be
described with either $P(r)=r^2$ or $P(r)=-1/r$ and this property allows very
fruitful comparisons.
\item $\lambda = 1$ because it corresponds to the funnel potential (Coulomb +
linear) which is widely used in hadron spectroscopy \cite{sema04}.
\end{itemize}

It could be also interesting to introduce potentials with $\lambda < -1$ but with
the restriction that it is repulsive at the origin (for instance, Van der Walls
forces or Lenhard-Jones type of potentials). In this case, the equation
determining $\nu_0$ is given by
\begin{equation}
\label{eq:detnu0potspec4} 
a - \nu_0= Z \nu_0^{\frac{1}{|\lambda| - 1}},
\end{equation}
which can be transformed with $1/(|\lambda| - 1) = p/q$ and $x=\nu_0^{1/q}$
into
\begin{equation}
\label{eq:detnu0potspec5} 
a - x^q = Z x^p.
\end{equation}
The set of all solvable potentials is provided with the list below:
\begin{equation}
\label{vallamb1srp}
\lambda = -5,\: -4, \: -3, \:-\frac{5}{2}, \:-\frac{7}{3},\: -2, \:-\frac{7}{4},
\: -\frac{5}{3}, \: -\frac{3}{2},  \:-\frac{4}{3},  \:-\frac{5}{4}.
\end{equation}
In this list, we will just consider the case:
\begin{itemize}
\item $\lambda = -2$ because the corresponding potential, known as the Kratzer
potential, exhibits its spectrum under an analytical form for all values of radial
quantum number $n$ and orbital quantum number $l$.
\end{itemize}

\subsection{Kratzer potential}
\label{subsec:Kratzer}

The Kratzer potential \cite{flu} is defined as
\begin{equation}
\label{eq:Kratz}
V(r)=\frac{a^2}{r^2}- \frac{2a}{r}.
\end{equation}
It presents some interest as a benchmark since it is one of the rare potentials
for which ones knows an exact analytical expression of the energies valid for
any $n$ and $l$ quantum numbers. Explicitly, one has
\begin{equation}
\label{eq:energKratex}
E(n,l) = - \frac{2ma^2}{\left [n+1/2 + \sqrt{(l+1/2)^2 + 2ma^2} \right ]^2}.
\end{equation}

Applying the AFM with $P(r)=-1/r$ to this potential leads to the following
equation giving the energies as a function of the auxiliary field $\nu$ (in this
section $N=N^{(\textrm{C})}=n+l+1$)
\begin{equation}
\label{eq:energKratanu}
E(\nu)=- \frac{m}{2 N^2}(2a - \nu)^2 - \frac{\nu^2}{4a^2}.
\end{equation}
The minimization with respect to $\nu$ provides the value
\begin{equation}
\label{eq:nu0Krat}
\nu_0=\frac{4ma^3}{2ma^2+N^2}
\end{equation}
which, inserted in (\ref{eq:energKratanu}), gives the desired result
\begin{equation}
\label{eq:energKrata}
E^{(\textrm{K})}(n,l)= - \frac{2ma^2}{\left [ 2ma^2 + (n+l+1)^2 \right ]}.
\end{equation}
Let us remark that the approximate value $E^{(\textrm{K})}$ presents the correct
asymptotic behaviour for large $n$ and for large $l$.
Just to have an idea of the quality of this approximation, let us calculate
the difference $\delta$ between the denominators of $E$ and 
$E^{(\textrm{K})}$. It is just a matter of simple algebra to find
\begin{equation}
\label{eq:difKrat}
\delta=(2n+1)(l+1/2) \left [ 1 - \sqrt{1+\frac{2ma^2}{(l+1/2)^2}} \right ].
\end{equation}
Consequently, for small intensity and/or mass, $ma^2 \ll 1$, or for large
angular momentum, $l \gg 1$, the approximate value tends to the exact one
and we have more explicitly
\begin{equation}
\label{eq:difKrata}
\delta \to - (2n+1) \frac{ma^2}{(l+1/2)}.
\end{equation}
This behaviour is easily understandable because, under those conditions, the
contribution due to $1/r$ is predominant as compared to the contribution of
$1/r^2$, and both expressions tends towards the same exact Coulomb result.

\subsection{Quadratic + centrifugal potential}
\label{subsec:quadcent}

We consider now the potential (for an attractive centrifugal potential, not
all values of $b$ are relevant \cite{case50})
\begin{equation}
\label{eq:quadcentpot}
V(r)= a r^2 \pm \frac{b}{r^2}.
\end{equation}
Incorporating the term $\pm b/r^2$ into the $l(l+1)/r^2$ term already present in
$\bm{p}^2$ allows to get the exact eigenvalue using the same kind of arguments
to those developed in the harmonic oscillator case \cite{flu}. Explicitly, we obtain
\begin{equation}
\label{eq:Eqpcent0}
E(n,l)=\sqrt{\frac{a}{2m}} \left[ 2(2n+1)+\sqrt{(2l+1)^2 \pm 8 m b}\right].
\end{equation}

Using the AFM with $P(r)=r^2$, the energies are given by (in this section
$N=N^{(\textrm{ho})}=2n+l+3/2$)
\begin{equation}
\label{eq:energqpcentnu}
E(\nu)=\sqrt{\frac{2}{m}} N (a \mp \nu)^{1/2} \pm 2 \sqrt{b} \; \nu^{1/2}.
\end{equation}
Setting $Y = 2mb/N^2$, the value $\nu_0$ that minimizes this energy comes 
from a first degree equation and reads
\begin{equation}
\label{eq:nu0qpcent}
\nu_0 = \frac{a Y}{1 \pm Y}.
\end{equation}
Substituting this value into the energy (\ref{eq:energqpcentnu}), 
one obtains a very simple expression for
the approximate energy
\begin{equation}
\label{eq:Eqpcent}
E^{(\textrm{qc})}(m,a,b;n,l)=2 \sqrt{\frac{a (N^2 \pm 2mb)}{2m}}.
\end{equation}

This quantity and the corresponding exact one depend on three parameters $m$,
$a$, $b$ but we know that the general scaling law properties allow us
to write them in a more pleasant form 
\begin{equation}
\label{eq:newfm1}
E(m,a,b;n,l) = \sqrt{\frac{a}{2m}} \; \epsilon(2mb;n,l),
\end{equation}
where $\epsilon(\beta;n,l)$ is an eigenvalue of the reduced
Schr\"{o}dinger equation for which the Hamiltonian depends now on a single
dimensionless parameter $\beta$
\begin{equation}
\label{eq:redschqpcent}
H=\bm{p}^2 + r^2 \pm \frac{\beta}{r^2}.
\end{equation}
The exact eigenvalues of this Hamiltonian are given by
\begin{equation}
\label{eq:engexqpcent}
\epsilon(\beta;n,l)=2(2n+1)+\sqrt{(2l+1)^2 \pm 4 \beta}.
\end{equation}
The approximate values immediately come from (\ref{eq:Eqpcent})
\begin{equation}
\label{eq:epsqpcent}
\epsilon^{(\textrm{qc})}(\beta;n,l) = 2 \sqrt{N^2 \pm \beta}.
\end{equation}
One can check that the relative error between $\epsilon$ and $\epsilon^{(\textrm{qc})}$
decreases as $l^{-3}$ for a fixed value of $n$ and 
decreases as $n^{-1}$ for a fixed value of $l$.
This behaviour is easily understandable because, for large values of the quantum numbers,
the contribution due to $r^2$ is predominant as compared to the contribution of
$1/r^2$ and both expressions tends towards the same exact harmonic oscillator result.

Let us assume that $\beta \ll 1$; a Taylor expansion truncated to first order
leads to
\begin{equation}
\label{eq:qpcentfo}
\epsilon^{(\textrm{qc})}(\beta;n,l) \approx 2N \pm \frac{\beta}{N}.
\end{equation}
In particular for $\beta=0$, one recovers the exact value $2 N$, as expected.
It is easy to check that this expression can also be
obtained by perturbation theory, as expected from section~\ref{subsec:relafmpert}.

\subsection{Some reduced polynomial equations}
\label{subsec:poleq}

The potentials that remain to be studied will need the solutions of cubic
and quartic equations. In order to simplify as much as possible the formulae,
we found interesting to put them in a form that makes the roots as simple as
possible. The corresponding notations will be used in the following.

\subsubsection{Cubic equation}
\label{subsubsec:Cubeq}

It is interesting to work with a cubic equation of the form
\begin{equation}
\label{eq:redcubeq1}
x^3 + 3x - 2Y =0.
\end{equation}
There exists only one positive root given analytically by
\begin{equation}
\label{eq:rootcubeq}
F(Y) = \left[Y + \sqrt{1 + Y^2} \right]^{1/3} - \left[Y + \sqrt{1 + Y^2}
\right]^{-1/3}.
\end{equation}
When $Y \ll 1$, one has $x$ close to 0 so that $x^3 \ll x$ and the behavior of the
root is simply
\begin{equation}
\label{eq:behYcsmall1}
F(Y) \approx \frac{2Y}{3} \quad \textrm{if} \quad Y \ll 1.
\end{equation}
When $Y \gg 1$, $x$ is large so that $3x$ is negligible with respect to $x^3$
and we have the following behaviour
\begin{equation}
\label{eq:behYcsmall2}
F(Y) \approx (2Y)^{1/3} \quad \textrm{if} \quad Y \gg 1.
\end{equation}

\subsubsection{Quartic equation}
\label{subsubsec:quarteq}

The quartic equation which gives the most pleasant form for the roots is
\begin{equation}
\label{eq:redcubeq2}
4 x^4 \pm 8x - 3Y =0.
\end{equation}
For each sign, there exists only one positive root given analytically by
\begin{equation}
\label{eq:rootquarteq}
G_{\pm}(Y) = \mp \frac{1}{2} \sqrt{V(Y)} + \frac{1}{2} \sqrt{ 4 (V(Y))^{-1/2}
- V(Y)},
\end{equation}
with
\begin{equation}
\label{eq:defVY}
V(Y)=\left(2 + \sqrt{4 + Y^3} \right)^{1/3} - Y \left(2 + \sqrt{4 + Y^3}
\right)^{-1/3}.
\end{equation}
When $Y \ll 1$, one has $x^4 \ll x$ and the behaviour of the roots is simply
\begin{equation}
\label{eq:behYqsmall1}
G_{+}(Y) \approx \frac{3Y}{8} \quad \textrm{if} \quad Y \ll 1,
\end{equation}
and
\begin{equation}
\label{eq:behYqsmall2}
G_{-}(Y) \approx 2^{1/3} + \frac{Y}{8} \quad \textrm{if} \quad Y \ll 1.
\end{equation}
When $Y \gg 1$, $x$ is large so that $8x$ is negligible with respect to $4 x^4$
and we have the following behaviour
\begin{equation}
\label{eq:behYcsmall3}
G_{\pm}(Y) \approx (3Y/4)^{1/4} \quad \textrm{if} \quad Y \gg 1.
\end{equation}

\subsection{Anharmonic potential}
\label{subsec:anarpot}

The potential under consideration reads
\begin{equation}
\label{eq:anharm}
V(r)= a r^2 + 2 b r.
\end{equation}
The interest for such a potential is discussed in section~\ref{sec:SolvablePotentials}.
Obviously, one must take $P(r)=r^2$ in the AFM and the energies to be
considered are given by (in this section $N=N^{(\textrm{ho})}=2n+l+3/2$)
\begin{equation}
\label{eq:energanarnu}
E(\nu)=\sqrt{\frac{2}{m}} N (a + \nu)^{1/2} + \frac{b^2}{\nu}.
\end{equation}
Let us introduce the parameter
\begin{equation}
\label{eq:Yanar}
Y = \frac{8}{3} a \left ( \frac{N^2}{m b^4} \right )^{1/3}
\end{equation}
and the new variable
\begin{equation}
x = \frac{3Y}{8a} \nu.
\end{equation}
The equation that leads to the minimization of the energy is then
\begin{equation}
\label{eq:eqredanar}
4 x^4 - 8 x - 3Y = 0.
\end{equation}
This is the reduced quartic equation studied in the preceding section.
The solution is given by $x_0(Y)=G_{-}(Y)$ (see (\ref{eq:rootquarteq})).
Substituting this value into the energy (\ref{eq:energanarnu}) and making
a little algebra leads to the desired approximate energy
\begin{equation}
\label{eq:Eanar}
E^{(\textrm{an})}(m,a,b;n,l)=\frac{3 b^2}{8a} Y \left ( G_{-}^2(Y) + \frac{1}{G_{-}(Y)}
\right ),
\end{equation}
with $Y$ given by (\ref{eq:Yanar}).

As in the previous case, this quantity and the corresponding exact one
depend on three parameters $m$, $a$, $b$ but the general scaling law allows us to
write them in terms of a reduced quantity depending on a single parameter $\beta$
\begin{equation}
\label{eq:newfm2}
E(m,a,b;n,l) = \sqrt{\frac{2a}{3m}} \; \epsilon \left (\frac{3b^2}{16}
\sqrt{\frac{3m}{2a^3}};n,l \right ),
\end{equation}
where $\epsilon(\beta;n,l)$ is an eigenvalue of the reduced
Schr\"{o}dinger equation for the Hamiltonian
\begin{equation}
\label{eq:redschanar}
H= \frac{\bm{p}^2}{4} + 3 r^2 + 8 \sqrt{\beta} r.
\end{equation}
The approximate value corresponding to this reduced equation follows from
(\ref{eq:Eanar}) 
\begin{equation}
\label{eq:epsanar}
\epsilon^{(\textrm{an})}(\beta;n,l) = 2 \, \beta \, Y \left ( G_{-}^2(Y) + 
\frac{1}{G_{-}(Y)} \right ), \quad Y = \left ( \frac{N}{\beta} \right )^{2/3}.
\end{equation}
The parameter $\beta$ could also be associated with the quadratic potential,
but this case less interesting is not considered here. 

Let us assume that $\beta \ll 1$; a Taylor expansion truncated to first order
leads to
\begin{equation}
\label{eq:anarfo}
\epsilon^{(\textrm{an})}(\beta;n,l) \approx \sqrt 3 N + 4 \sqrt{\frac{2 \beta \, N}
{\sqrt 3}}.
\end{equation}
In particular for $\beta=0$, one recovers the exact value $\sqrt 3 N$, as
it should be. This result comes also from the perturbation theory.

The limit $\beta\to\infty$ is not physically relevant, but it is interesting to
consider it in order to check the formula. In this limit, we find
\begin{equation}
\label{eq:anbetinf}
\epsilon^{(\textrm{an})}(\beta;n,l) = 3(4 \beta N^2)^{1/3}+\Or\left(
\beta^{-1/3} \right).
\end{equation}
The dominant term is the result expected for a pure linear potential, as given by
(\ref{eq:eigenerplpot}).

\subsection{Quadratic + Coulomb potential}
\label{subsec:quadcoul}

In this section, we study the quadratic + Coulomb potential defined as
\begin{equation}
\label{eq:quadcoulpot}
V(r)= a r^2 - \frac{b}{r}.
\end{equation}
The interest for such a potential is discussed in section~\ref{sec:SolvablePotentials}.
Let us illustrate the AFM method with the option $P(r)=r^2$ so that 
$N=N^{(\textrm{ho})}=2n+l+3/2$. The energies depending on the auxiliary
field $\nu$ are given in this case by
\begin{equation}
\label{eq:energqpcoulnu}
E(\nu)=\sqrt{\frac{2}{m}} N (a + \nu)^{1/2} -3 \left ( \frac{b^2 \nu}{4}
\right )^{1/3}.
\end{equation}
Let us introduce the parameter
\begin{equation}
\label{eq:Yqpcoul}
Y = \frac{8 N^2}{3m} \left ( \frac{4a}{b^4} \right )^{1/3}
\end{equation}
and the new variable
\begin{equation}
x = \left ( \frac{2a}{\nu} \right )^{1/3}.
\end{equation}
The equation that leads to the minimization of the energy is then
\begin{equation}
\label{eq:eqredqpcoul}
4 x^4 + 8 x - 3Y = 0.
\end{equation}
This reduced quartic equation was studied previously. The solution is given
by $x_0(Y)=G_{+}(Y)$ (see (\ref{eq:rootquarteq})).
Substituting this value into the energy (\ref{eq:energqpcoulnu}), one is led,
after some manipulations, to the desired approximate energy
\begin{equation}
\label{eq:Eqpcoul}
E^{(\textrm{qC})}(m,a,b;n,l)=\frac{3}{4} \left ( \frac{a b^2}{2} \right )^{1/3}
\left [ \frac{Y}{G_{+}^2(Y)} - \frac{4}{G_{+}(Y)} \right ],
\end{equation}
with $Y$ given by (\ref{eq:Yqpcoul}).

One can check that, starting with $P(r)=-1/r$ and performing the same kind of
algebra, the approximate energy is given by exactly the same equation as
(\ref{eq:Eqpcoul}), but this time with the $Y$ parameter given by
(\ref{eq:Yqpcoul}) in which $N^{(\textrm{ho})}$ is replaced by $N^{(\textrm{C})}$.
This property is a very nice check of the general prescription demonstrated
in section \ref{subsec:switchlmu}.

The quantity (\ref{eq:Eqpcoul}) and the corresponding exact one depend again on three parameters
$m$, $a$, $b$ but the general scaling law allows us to write them in terms of a
reduced quantity depending on a single dimensionless parameter $\beta$. One can
imagine two formulations depending on whether $\beta$ is part of the quadratic
contribution or of the Coulomb contribution:
\begin{eqnarray}
\label{eq:newfm3}
E(m,a,b;n,l)&=&4 \sqrt{\frac{2a}{3m}} \; \epsilon \left (\frac{1}{4}
\left ( \frac{54 m^3 b^4}{a} \right )^{1/6};n,l \right ), \\
\label{eq:newfm4}
E(m,a,b;n,l)&=&\frac{3 m b^2}{16} \; \eta \left ( 4 \left (\frac{a}{54 m^3 b^4}
\right )^{1/6};n,l \right ).
\end{eqnarray}
The $\epsilon$ and $\eta$ energies are the eigenvalues of the reduced
Schr\"{o}dinger equations for the respective Hamiltonians $H_\epsilon$ et
$H_\eta$:
\begin{eqnarray}
\label{eq:redschqpcoul1}
H_\epsilon&=&\frac{3 \bm{p}^2}{16} +\frac{r^2}{4} - \frac{\beta^{3/2}}{r}, \\
\label{eq:redschqpcoul2}
H_\eta&=&\frac{3 \bm{p}^2}{16} - \frac{\sqrt 2}{r} + {\beta '}^{6} r^2.
\end{eqnarray}
The approximate values corresponding to these reduced Hamiltonians follow from
(\ref{eq:Eqpcoul}):
\begin{eqnarray}
\label{eq:epsqpcoul}
\epsilon^{(\textrm{qC})}(\beta;n,l)&=&\frac{3 \beta}{8} 
\left [ \frac{Y}{G_{+}^2(Y)} - \frac{4}{G_{+}(Y)} \right ], \quad Y = 
\left ( \frac{N}{\beta} \right )^2, \\
\label{eq:etaqpcoul}
\eta^{(\textrm{qC})}(\beta ';n,l)&=&\frac{3 {\beta '}^2}{4} 
\left [ \frac{Y}{G_{+}^2(Y)} - \frac{4}{G_{+}(Y)} \right ], \quad Y = 
\left ( N \beta' \right )^2.
\end{eqnarray}

Let us assume that, $\beta \ll 1$, that is $V(r) \ll P(r)$:
\begin{itemize}
\item A Taylor expansion truncated to first order for the formulation based
on the $\epsilon$ form
leads to
\begin{equation}
\label{eq:qpcoulfo1}
\epsilon^{(\textrm{qC})}(\beta;n,l) \approx \frac{\sqrt 3}{4} N - \sqrt{\frac{2 \beta^3}
{N \sqrt 3}}.
\end{equation}
In particular for $\beta=0$, one recovers the exact value $\sqrt 3 N/4$
if $N=N^{(\textrm{ho})}$. 
\item The Taylor expansion for the formulation based on the $\eta$ form gives
\begin{equation}
\label{eq:qpcoulfo2}
\eta^{(\textrm{qC})}(\beta ';n,l) \approx - \frac{8}{3 N^2} +
\frac{9 {\beta '}^6 N^4}{128}.
\end{equation}
In particular for $\beta '=0$, one recovers the exact value $- 8/(3 N^2)$ if
$N=N^{(\textrm{C})}$. 
\end{itemize}
In all cases, the approximate formulae resulting from AFM agree with the result
of perturbation theory.

The limit $\beta\to\infty$ is interesting to consider in order to check the formulae: 
\begin{itemize}
\item 
\begin{equation}
\label{eq:qcbetinf1}
\epsilon^{(\textrm{qC})}(\beta;n,l) = -\frac{4\beta^3}{3 N^2}+\Or\left(
\beta^{-2} \right).
\end{equation}
If $N=N^{(\textrm{C})}$, the dominant term is the exact result for a Coulomb
potential.
\item 
\begin{equation}
\label{eq:qcbetinf2}
\eta^{(\textrm{qC})}(\beta';n,l) = \frac{\sqrt{3}}{2}{\beta'}^3 N+
\Or\left( {\beta'}^{3/2} \right).
\end{equation}
If $N=N^{(\textrm{ho})}$, the dominant term is the exact result for a
quadratic potential.
\end{itemize}

Let us emphasize the point that both (\ref{eq:redschqpcoul1}) and
(\ref{eq:redschqpcoul2}) can be related with the scaling laws developed in
section \ref{subsec:scallaw}. As a consequence, it can be shown that both the
exact eigenvalues and the AFM approximate ones fulfill the relation
\begin{equation}
\label{eq:linkepseta}
\epsilon(\beta;n,l)=\frac{\beta^3}{2} \eta(1/\beta;n,l). 
\end{equation}

\subsection{Funnel potential}
\label{subsec:funnel}

In this section, we are concerned with the funnel potential defined as
\begin{equation}
\label{eq:funpot}
V(r)= a r - \frac{b}{r}.
\end{equation}
This potential is particularly important and its interest is discussed 
in section~\ref{sec:SolvablePotentials}. Let us recall that it is widely used
in hadronic spectroscopy and corresponds to a linear confinement coupled to a
short range Coulomb contribution \cite{sema04}. Finding approximate analytical
values for the energies corresponding to this potential is thus a very
interesting question. To our knowledge such formulae are not proposed in
literature.

Naturally, one must apply the AFM method with  $P(r)=-1/r$ so that 
$N=N^{(\textrm{C})}=n+l+1$. The energies depending on the auxiliary
field $\nu$ can be calculated following the prescription detailed in section
\ref{subsec:Extens}. Explicitly, one finds
\begin{equation}
\label{eq:energfunnu}
E(\nu)=- \frac{m(b+\nu)^2}{2 N^2} + 2 \sqrt a \nu^{1/2}.
\end{equation}
Let us introduce the parameter
\begin{equation}
\label{eq:Yfun}
Y = \frac{3}{2} N^2 \sqrt{ \frac{3a}{m^2 b^3}}
\end{equation}
and the new variable
\begin{equation}
x = \sqrt{\frac{3\nu}{b}}.
\end{equation}
The equation that must be solved to minimize the energy is
\begin{equation}
\label{eq:eqredfun}
x^3 + 3 x - 2Y = 0.
\end{equation}
This reduced cubic equation was studied previously. The solution is given
by $x_0(Y)=F(Y)$ (see (\ref{eq:rootcubeq})).
Inserting this value into the energy (\ref{eq:energfunnu}), one finds the
expression of the approximate energy
\begin{equation}
\label{eq:Efun}
E^{(\textrm{f})}(m,a,b;n,l) 
= \sqrt{3 a b} \left [ \frac{Y}{F^2(Y)} - \frac{2}{F(Y)} \right ],
\end{equation}
with $Y$ given by (\ref{eq:Yfun}).
Let us mention that another form of this equation can be found thanks to
the following relation
\begin{equation}
\label{eq:Efun2}
\frac{Y}{F^2(Y)} - \frac{2}{F(Y)} 
= \sinh \theta -\frac{1}{4 \sinh \theta},
\end{equation}
with the change of variables $Y=\sinh (3 \theta)$.

The quantity (\ref{eq:Efun}) and the corresponding exact one depend again on three parameters
$m$, $a$, $b$ but the general scaling law allows us to write them in terms of a
reduced quantity depending on a single dimensionless parameter $\beta$. 
In hadronic physics, the dominant interaction between a quark and an antiquark is a 
confining linear potential \cite{sema04}. So, 
although the linear potential has no analytical exact solution for all values of the 
quantum numbers, it is also interesting to consider two formulations depending on 
whether $\beta$ is part of the linear contribution or of the Coulomb contribution:
\begin{eqnarray}
\label{eq:newfm5}
E(m,a,b;n,l)&=&3 \left ( \frac{a^2}{2m} \right )^{1/3} \epsilon \left (
\left ( \frac{4 m^2 b^3}{27 a} \right )^{1/4};n,l \right ), \\
\label{eq:newfm6}
E(m,a,b;n,l)&=&\frac{2 m b^2}{3^{5/3}} \; \eta \left (\left (\frac{27 a}
{4 m^2 b^3} \right )^{1/4};n,l \right ).
\end{eqnarray}
The $\epsilon$ and $\eta$ energies are the eigenvalues of the reduced
Schr\"{o}dinger equations for the respective Hamiltonians $H_\epsilon$ et
$H_\eta$:
\begin{eqnarray}
\label{eq:redschfun1}
H_\epsilon&=&\frac{\bm{p}^2}{3} +\frac{r}{3} - \frac{\beta^{4/3}}{r}, \\
\label{eq:redschfun2}
H_\eta&=&\frac{\bm{p}^2}{3} - \frac{3^{1/3}}{r} + {\beta '}^{4} r.
\end{eqnarray}
The approximate values corresponding to these reduced Hamiltonian follow from
(\ref{eq:Efun}):
\begin{eqnarray}
\label{eq:epsfun}
\epsilon^{(\textrm{f})}(\beta;n,l)&=&\beta^{2/3} 
\left [ \frac{Y}{F^2(Y)} - \frac{2}{F(Y)} \right ], 
\quad Y = \left ( \frac{N}{\beta} \right )^2, \\
\label{eq:etafun}
\eta^{(\textrm{f})}(\beta ';n,l)&=&3^{2/3} {\beta '}^2 
\left [ \frac{Y}{F^2(Y)} - \frac{2}{F(Y)} \right ], \quad Y = \left ( N
\beta ' \right )^2.
\end{eqnarray}

Let us assume that $\beta \ll 1$, that is $V(r) \ll P(r)$:
\begin{itemize}
\item The Taylor expansion for the formulation based on the $\epsilon$ form gives
\begin{equation}
\label{eq:funfo1}
\epsilon^{(\textrm{f})}(\beta;n,l) \approx \frac{N^{2/3}}{2^{2/3}}  - \left (
\frac{\beta^4}{2 N^2} \right )^{1/3}.
\end{equation}
In particular for $\beta=0$, one recovers the value expected for a pure linear
potential, as given by (\ref{eq:eigenerplpot}).
\item The Taylor expansion for the formulation based on the $\eta$ form gives
\begin{equation}
\label{eq:funfo2}
\eta^{(\textrm{f})}(\beta ';n,l) \approx - \frac{3^{5/3}}{4 N^2} + \frac{2 N^2
{\beta '}^4} {3^{4/3}}.
\end{equation}
In particular for $\beta'=0$, one recovers the exact value $- 3^{5/3}/(4 N^2)$.
\end{itemize}
In all cases, the approximate formulae resulting from AFM agree with the result
of perturbation theory.

The limit $\beta\to\infty$ is interesting to consider in order to check the formulae: 
\begin{itemize}
\item 
\begin{equation}
\label{eq:funbetinf1}
\epsilon^{(\textrm{f})}(\beta;n,l) = -\frac{3\beta^{8/3}}{4 N^2}+\Or\left( 
\beta^{-4/3} \right).
\end{equation}
The dominant term is the exact result for a Coulomb
potential.
\item 
\begin{equation}
\label{eq:funbetinf2}
\eta^{(\textrm{f})}(\beta';n,l) = \left(\frac{3}{2}{\beta'}^4 N\right)^{2/3} +
\Or\left( {\beta'}^{4/3} \right).
\end{equation}
The dominant term is the result expected for a pure linear potential, as given by
(\ref{eq:eigenerplpot}).
\end{itemize}

Here again, both (\ref{eq:redschfun1}) and (\ref{eq:redschfun2}) can
be related with the scaling laws. As a consequence, it can be shown that both
exact eigenvalues and AFM approximate ones fulfill the relation
\begin{equation}
\label{eq:funfo3}
\epsilon(\beta;n,l) = \frac{\beta^{8/3}}{3^{2/3}}  \eta (1/\beta;n,l).
\end{equation}

\section{Comparison with numerical results}
\label{sec:numres}

In the previous section, approximate analytical forms for eigenvalues of several
Hamiltonians were found. The formulae depend on the quantum numbers $n$ and $l$
through a factor $N$. This number could be taken as $N^{(\textrm{ho})}$, 
$N^{(\textrm{C})}$ or even $N_\lambda$ if the potential $P(r)$ chosen is the
power-law potential $P^{(\lambda)}(r)$ (see section~\ref{subsec:switchlmu}). If we
look, for instance, at the Hamiltonian~(\ref{eq:redschqpcoul1}), it is clear that
it reduces to a harmonic oscillator when $\beta=0$. In this case, the choice
$N=N^{(\textrm{ho})}$ gives the exact result. When $\beta \to \infty$, the Coulomb
part dominates and the choice $N=N^{(\textrm{C})}$ is expected to yield the exact
result asymptotically. 

All dimensionless Hamiltonians considered above depend on a parameter $\beta$. The
variation of the eigenvalues being smooth for the variation of $\beta$, we can
assume that the number $N$, giving the optimal values for a selected set of these
eigenvalues, is also a smooth function of $\beta$. From considerations above, the
functional form 
\begin{equation}
\label{Nbeta}
N(\beta)= b(\beta) n + l + c(\beta) 
\end{equation}
seems reasonable. If, in the limits $\beta\to 0$ and $\beta \to \infty$, the
Hamiltonian reduces to a known form, the values $N(0)$ and $N(\infty)$ can be computed.
But for finite values of $\beta$, we cannot predict the correct behaviour. It is then
necessary to focus our attention on numerical solutions. 
 
Very accurate eigenvalues $\epsilon_{\textrm{num}}(\beta;n,\ell)$ for Hamiltonians
defined above can be obtained numerically with the Lagrange mesh method \cite{lag}.
It is very accurate and easy to implement. In order to find the best possible
values for coefficients $d(\beta)$ ($d$ stands for $b$ or $c$), we will use the
following measure
\begin{equation}
\label{chi2}
\chi(\beta)=\frac{1}{16}\sum_{n=0}^{3}\sum_{\ell=0}^{3} \left( \epsilon_{\textrm{num}}
(\beta;n,\ell) - \epsilon_{\textrm{app}}(\beta;n,\ell) \right)^2,
\end{equation}
where $\epsilon_{\textrm{app}}(\beta;n,\ell)$ are values obtained from our
approximate formulae.
Other choices are possible but we find this one very convenient. 
The analytical form $\epsilon_{\textrm{app}}$ depends on $N(\beta)$ which depends
on coefficients $d$. 
For each value of $\beta$, optimal values for the $d$ coefficients, $d_{\textrm{min}}
(\beta)$, can be determined by minimizing $\chi(\beta)$. Then, with a set
\{$d_{\textrm{min}}(\beta)$\} for a given set \{$\beta$\}, a functional form
$d_{\textrm{fit}}(\beta)$ can be fitted with the following measure
\begin{equation}
\label{chi2bis}
\chi(d)=\sum_{\{\beta\}} \left( d_{\textrm{min}}
(\beta) - d_{\textrm{fit}}(\beta) \right)^2.
\end{equation}
Again, other choices are possible but we find this one very convenient. We will now
try to determine the best form of coefficients $d(\beta)$ for some of the potentials
studied above. 

\subsection{Improvement for anharmonic potential}
\label{subsec:Ian}

\begin{figure}[ht]
\begin{center}
\includegraphics*[width=6.4cm]{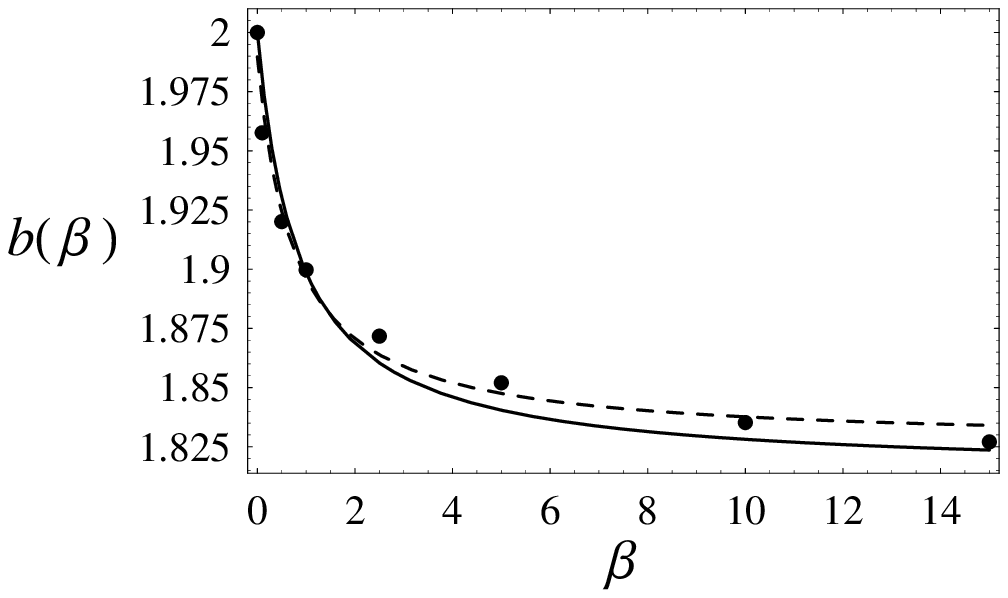}
\includegraphics*[width=6.4cm]{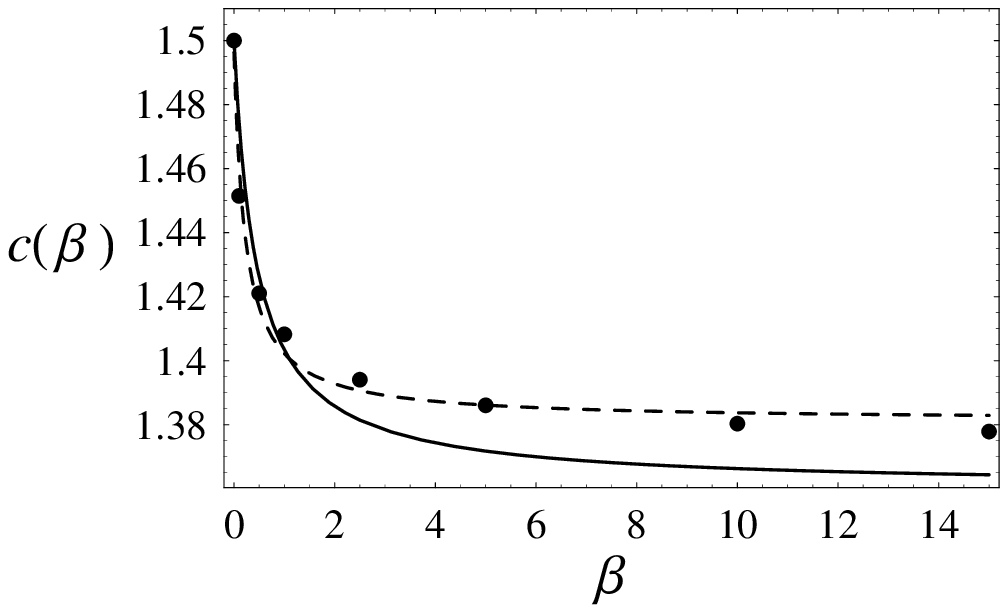}
\caption{\label{fig:an}Best values of the coefficients $b(\beta)$ and $c(\beta)$
to parametrize the eigenvalues of Hamiltonian~(\ref{eq:angraph}): numerical fit with
(\ref{chi2}) [dots]; functions (\ref{eq:bcan}) with set 1 of parameters from
table~\ref{tab:an} [solid line]; idem with set 2 [dashed line].} 
\end{center}
\end{figure}

The dimensionless Hamiltonian proposed here for the anharmonic potential is 
\begin{equation}
\label{eq:angraph}
H= \frac{\bm{p}^2}{4} + 3 r^2 + 8 \sqrt{\beta} r.
\end{equation}
Approximate eigenvalues are given by (\ref{eq:epsanar}).
With (\ref{Nbeta}), the exact result will be obtained for $b(0)=2$ and $c(0)=3/2$.
When $\beta \to \infty$, the linear part dominates and, from results obtained in SSB,
we could expect that $b(\infty)\approx \pi/\sqrt{3}\approx 1.814$ and $c(\infty)
\approx \sqrt{3}\pi/4\approx 1.360$. By minimizing our measure $\chi(\beta)$, we
found the optimal values of $b(\beta)$ and $c(\beta)$ for several finite values of
$\beta$. The results are plotted with dots on figure~\ref{fig:an}. One can clearly
see the smooth transition between two domains for zero and infinite values of
$\beta$. This corresponds to the transition between two potentials not too different,
a linear one and a quadratic one. 

We tried to fit the numerical points with various functions and found that the best
result is obtained for sections of hyperbola, both for $b(\beta)$ and $c(\beta)$. The
parameterization retained is 
\begin{equation}
\label{eq:bcan}
b(\beta)= \frac{p_1 \beta+p_2}{\beta+p_3}, \quad 
c(\beta)= \frac{q_1 \beta+q_2}{\beta+q_3}.
\end{equation}
With this choice, $b(0)=p_2/p_3$, $b(\infty)=p_1$, $c(0)=q_2/q_3$ and $c(\infty)=q_1$.
Two different fits are presented in table~\ref{tab:an}. For set 1, only one parameter
is free for each coefficient and the following constraints are imposed: $b(0)=2$,
$b(\infty)= \pi/\sqrt{3}$, $c(0)=3/2$, $c(\infty)=\sqrt{3}\pi/4$.
For set 2, all parameters are free; the results found by minimization give: $b(0)
\approx 1.990$ close to 2, $b(\infty)= 1.826$ close to 1.814 ($\pi/\sqrt{3}$),
$c(0)\approx 1.496$ close to 1.5, $c(\infty)=1.381$ close to 1.360 ($\sqrt{3}\pi/4$). 

\begin{table}[htb]
\caption{\label{tab:an}Values of parameters $p_i$ and $q_i$ for (\ref{eq:bcan}).
Fixed parameters are marked by a *.}
\begin{indented}
\item[]\begin{tabular}{@{}ccccccc}
\br
 & $p_1$ & $p_2$ & $p_3$ & $q_1$ & $q_2$ & $q_3$  \\
\mr
Set 1 & $\pi/\sqrt{3}$* & $2 p_3$* & 0.835 & $\sqrt{3}\pi/4$* & $3 q_3/2$* & 0.445 \\
Set 2 & 1.826 & 1.485 & 0.747 & 1.381 & 0.333 & 0.222 \\
\br
\end{tabular}
\end{indented}
\end{table}

The quality of the fits can be appraised by examining the values of $\chi(\beta)$
shown in table~\ref{tab:an2}. It is clear that allowing a $\beta$-dependence for
the coefficients $b$ and $c$ improves greatly the approximate eigenvalues. Let
us remark that the fit with three parameters is only slightly better than the fit
with only one parameter, and that $\chi(\beta)$ increases with $\beta$ for the
choice $N=N^{(\textrm{ho})}$, as expected since the potential deviates more and
more from a pure quadratic one.

\begin{table}[htb]
\caption{\label{tab:an2}Values of $\chi(\beta)$ for eigenvalues of 
Hamiltonian~(\ref{eq:angraph})
as a function of $\beta$, for various parameterization of
$N=b(\beta) n + l + c(\beta)$ (see table~\ref{tab:an}). $N^{(\textrm{ho})}$
and set 1 give the exact result for $\beta=0$.}
\begin{indented}
\item[]\begin{tabular}{@{}cccc}
\br
$\beta$ & $N^{(\textrm{ho})}$ & Set 1 & Set 2 \\
\mr
0.1 & $5.9\ 10^{-2}$ & $6.2\ 10^{-3}$ & $4.0\ 10^{-3}$ \\
1   & $0.46$         & $2.7\ 10^{-3}$ & $1.9\ 10^{-3}$ \\
10  & $2.8$          & $1.8\ 10^{-2}$ & $5.4\ 10^{-3}$ \\
\br
\end{tabular}
\end{indented}
\end{table}

\subsection{Improvement for quadratic + Coulomb potential}
\label{subsec:Iqc}

\begin{figure}[ht]
\begin{center}
\includegraphics*[width=6.4cm]{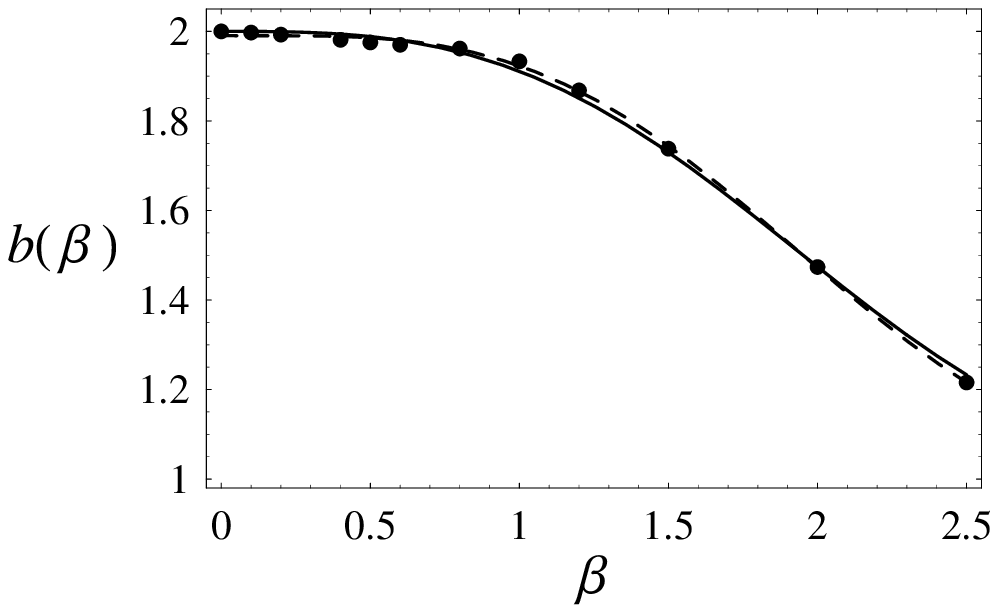}
\includegraphics*[width=6.4cm]{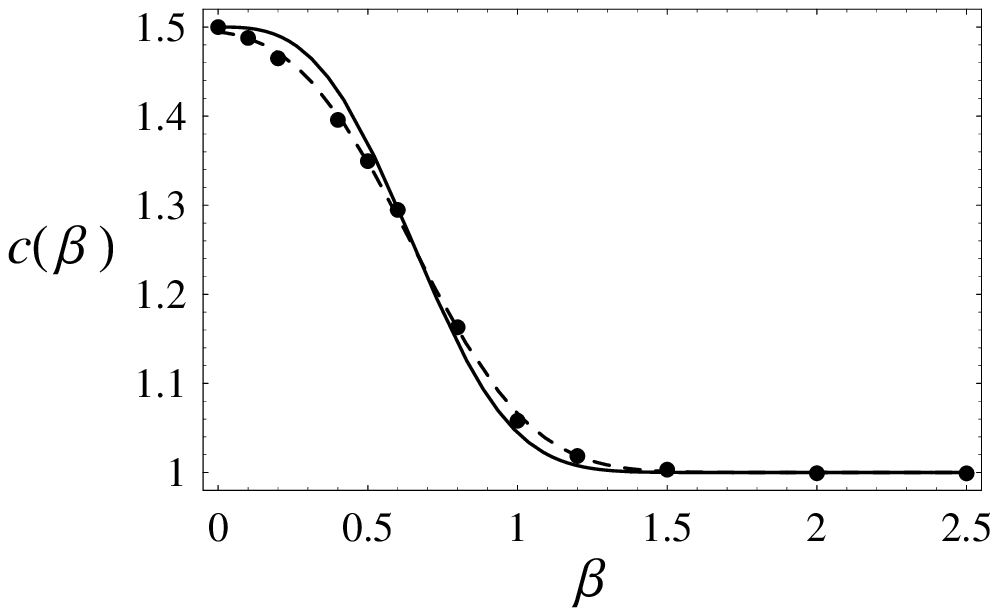}
\caption{\label{fig:qc}Best values of the coefficients $b(\beta)$ and $c(\beta)$
to parametrize the eigenvalues of Hamiltonian~(\ref{eq:qcgraph}): numerical fit
with (\ref{chi2}) [dots]; functions (\ref{eq:bcqc}) with set 1 of parameters from
table~\ref{tab:qc} [solid line]; idem with set 2 [dashed line].} 
\end{center}
\end{figure}

To study the case of the quadratic + Coulomb potential, we choose the following
dimensionless Hamiltonian 
\begin{equation}
\label{eq:qcgraph}
H=\frac{3 \bm{p}^2}{16} +\frac{r^2}{4} - \frac{\beta^{3/2}}{r}.
\end{equation}
Approximate eigenvalues are given by (\ref{eq:epsqpcoul}).
The exact result will be obtained for $b(0)=2$ and $c(0)=3/2$. When $\beta \to
\infty$, the Coulomb part dominates and we could expect that $b(\infty)=1$ and
$c(\infty)=1$. By minimizing our measure $\chi(\beta)$, we found the optimal
values of $b(\beta)$ and $c(\beta)$ for several finite values of $\beta$. The
results are plotted with dots on figure~\ref{fig:qc}. One can clearly see the
smooth transition between the harmonic oscillator region near zero and the
asymptotic Coulomb region. In this case, the transition occurs between two very
different potentials: $r^2$ and $-1/r$. It is clear that the asymptotic region
is more rapidly reached for $c(\beta)$ than for $b(\beta)$. When the quantum
numbers $n$ and $l$ increase, the size of the eigenfunctions grows and the states
become less sensitive to the Coulomb part. So, we can understand that the influence
of $\beta$ on the coefficient $b(\beta)$ is less significant than on $c(\beta)$.

We tried to fit the numerical points with various functions and found that the
best result is obtained for exponential functions with a cubic argument, both for
$b(\beta)$ and $c(\beta)$. The parameterization retained is 
\begin{equation}
\fl
\label{eq:bcqc}
b(\beta)= 1+p_1 \exp\left( -p_2(\beta-p_3)^3\right), \quad
c(\beta)= 1+q_1 \exp\left( -q_2(\beta-q_3)^3\right).
\end{equation}
With this choice, $b(\infty)=1$ and $c(\infty)=1$.
Two different fits are presented in table~\ref{tab:qc}. For set 1, only one
parameter is free for each coefficient and the following constraints are imposed:
$b(0)=2$ and $c(0)=3/2$. For set 2, all parameters are free; the results found
by minimization give: $b(0)\approx 1.991$ close to 2 and $c(0)\approx 1.495$
close to 1.5.

\begin{table}[htb]
\caption{\label{tab:qc}Values of parameters $p_i$ and $q_i$ for (\ref{eq:bcqc}).
Fixed parameters are marked by a *.}
\begin{indented}
\item[]\begin{tabular}{@{}ccccccc}
\br
 & $p_1$ & $p_2$ & $p_3$ & $q_1$ & $q_2$ & $q_3$  \\
\mr
Set 1 & 1* & 0.093 & 0* & $1/2$* & 2.414 & 0* \\
Set 2 & 0.990 & 0.119 & 0.161 & 0.496 & 1.373 & $-0.136$ \\
\br
\end{tabular}
\end{indented}
\end{table}

The quality of the fits can be appraised by examining the values of $\chi(\beta)$
shown in table~\ref{tab:qc2}. Again, allowing a $\beta$-dependence for the
coefficients $b$ and $c$ improves greatly the approximate eigenvalues. The fits,
with one parameter and three parameters, give nearly the same result. 
$\chi(\beta)$ increases with $\beta$ for the choice $N=N^{(\textrm{ho})}$, as
expected since the potential deviates more and more from a pure quadratic one.
On the contrary, $\chi(\beta)$ decreases with $\beta$ for the choice $N=
N^{(\textrm{C})}$, as expected since the potential comes closer to a pure Coulomb
one. The values of $\chi(\beta)$ are large because the considered values of
$\beta$ are not in the asymptotic region of $b(\beta)$, as one can see on 
figure~\ref{fig:qc}. It has been checked that $\chi(\beta)\to 0$ also for sets 1 and 2
for large values of $\beta$.

\begin{table}[htb]
\caption{\label{tab:qc2}Values of $\chi(\beta)$ for eigenvalues of
Hamiltonian~(\ref{eq:qcgraph}) as a function of $\beta$, for various
parameterizations of $N=b(\beta) n + l + c(\beta)$ (see table~\ref{tab:qc}).
$N^{(\textrm{ho})}$ gives the exact result for $\beta=0$, $N^{(\textrm{C})}$ gives
the exact result for $\beta=\infty$, set 1 gives the exact result for both
$\beta=0$ and $\beta=\infty$, and set 2 gives the exact result for $\beta=\infty$.}
\begin{indented}
\item[]\begin{tabular}{@{}ccccc}
\br
$\beta$ & $N^{(\textrm{ho})}$ & $N^{(\textrm{C})}$ & Set 1 & Set 2 \\
\mr
0.5 & $9.6\ 10^{-3}$ & 0.90 & $2.6\ 10^{-3}$ & $2.3\ 10^{-3}$ \\
1   & $0.11$         & 0.75 & $1.9\ 10^{-2}$ & $1.8\ 10^{-2}$ \\
2   & $3.1$          & 0.45 & $0.12$         & $0.12$ \\
\br
\end{tabular}
\end{indented}
\end{table}

\subsection{Improvement for funnel potential}
\label{subsec:Ifunnel}

\begin{figure}[ht]
\begin{center}
\includegraphics*[width=6.4cm]{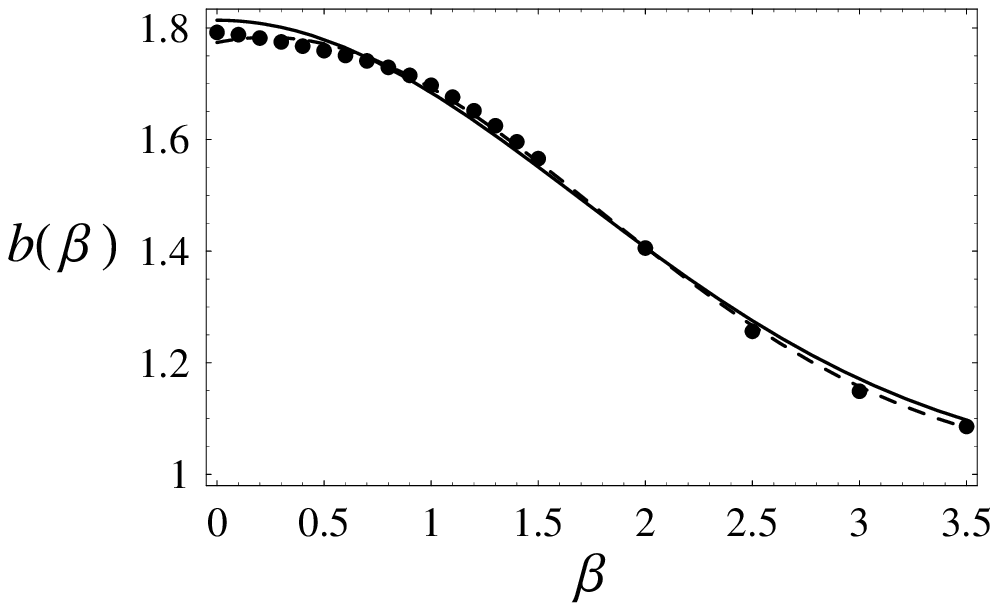}
\includegraphics*[width=6.4cm]{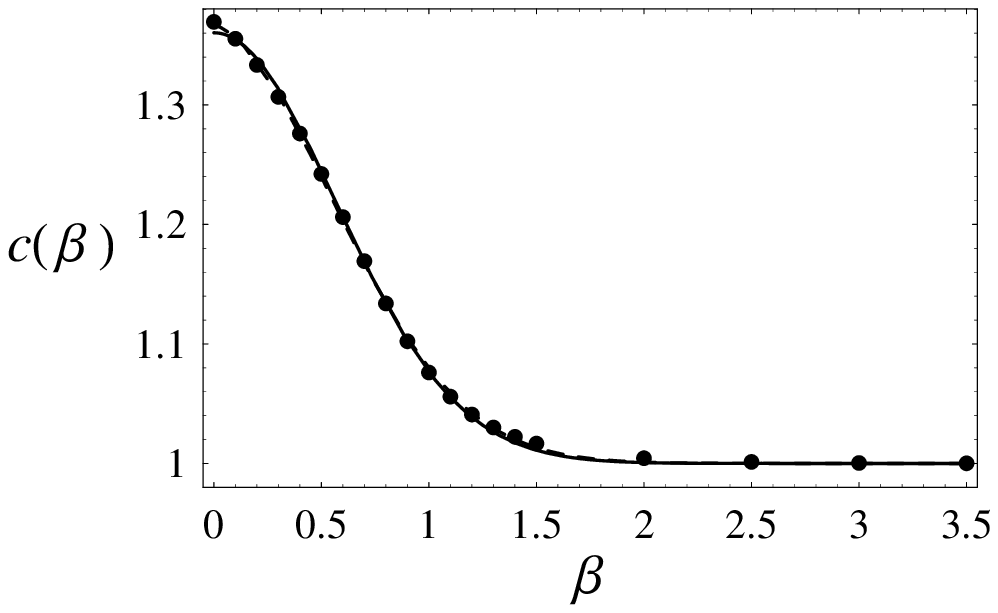}
\caption{\label{fig:fun}Best values of the coefficients $b(\beta)$ and
$c(\beta)$ to parametrize the eigenvalues of Hamiltonian~(\ref{eq:fungraph}):
numerical fit with (\ref{chi2}) [dots]; functions (\ref{eq:bcfun}) with
set 1 of parameters from table~\ref{tab:fun} [solid line]; idem with set
2 [dashed line].} 
\end{center}
\end{figure}

For the case of the funnel potential, we choose the following dimensionless
Hamiltonian 
\begin{equation}
\label{eq:fungraph}
H=\frac{\bm{p}^2}{3} +\frac{r}{3} - \frac{\beta^{4/3}}{r}.
\end{equation}
Approximate eigenvalues are given by (\ref{eq:epsfun}).
This choice is motived by three reasons:
\begin{itemize}
\item For hadronic problems \cite{sema04}, physical values of $\beta$
varies from 0 to about 1.5;
\item A comparison is possible between Hamiltonians~(\ref{eq:fungraph})
and (\ref{eq:qcgraph}) which differ by the `confinement' part.
\item Contrary to previous cases, $\beta=0$ does not correspond to a
full analytical case;
\end{itemize}
For $\beta=0$, the potential is pure linear one. So, from the study
performed in SSB, we can expect that $b(0)= \pi/\sqrt{3}\approx 1.814$ and
$c(0)= \sqrt{3}\pi/4\approx 1.360$ is a good choice. When $\beta \to \infty$,
the Coulomb part dominates and we could expect that $b(\infty)=1$ and
$c(\infty)=1$. By minimizing our measure $\chi(\beta)$, we found the optimal
values of $b(\beta)$ and $c(\beta)$ for several finite values of $\beta$. The
results are plotted with dots on figure~\ref{fig:fun}. One can see a smooth
transition similar to the previous case. Again, for probably the same reason,
the asymptotic region is more rapidly reached for $c(\beta)$ than for
$b(\beta)$. 

We tried to fit the numerical points with various functions and found that
the best result is obtained for a Gaussian function both for $b(\beta)$ and
$c(\beta)$. The parameterization retained is 
\begin{equation}
\fl
\label{eq:bcfun}
b(\beta)= 1+p_1 \exp\left( -p_2^2(\beta-p_3)^2\right), \quad
c(\beta)= 1+q_1 \exp\left( -q_2^2(\beta-q_3)^2\right).
\end{equation}
Let us note the difference with the previous case, for which the argument of
the exponential was cubic in $\beta$ and not quadratic. With this choice,
$b(\infty)=1$ and $c(\infty)=1$. Two different fits are presented in 
table~\ref{tab:fun}. For set 1, only one parameter is free for each coefficient
and the following constraints are imposed: $b(0)=\pi/\sqrt{3}$ and
$c(0)=\sqrt{3}\pi/4$. For set 2, all parameters are free; the results found
by minimization give: $b(0)\approx 1.774$ close to $\pi/\sqrt{3}$ (1.814) and
$c(0)\approx 1.367$ close to $\sqrt{3}\pi/4$ (1.360). 

\begin{table}[htb]
\caption{\label{tab:fun}Values of parameters $p_i$ and $q_i$ for
(\ref{eq:bcfun}). Fixed parameters are marked by a *.}
\begin{indented}
\item[]\begin{tabular}{@{}ccccccc}
\br
 & $p_1$ & $p_2$ & $p_3$ & $q_1$ & $q_2$ & $q_3$  \\
\mr
Set 1 & $\frac{\pi}{\sqrt{3}}-1$* & 0.416 & 0* & $\frac{\sqrt{3}\pi}
{4}-1$* & 1.245 & 0* \\
Set 2 & 0.783 & 0.459 & 0.237 & 0.369 & 1.168 & $-0.062$ \\
\br
\end{tabular}
\end{indented}
\end{table}

The quality of the fits can be appraised by examining the values of
$\chi(\beta)$ shown in table~\ref{tab:fun2} and the values of 
approximate results compared with exact ones presented 
in table~\ref{tab:fun3}. As in the two previous
cases, allowing a $\beta$-dependence for the coefficients $b$ and $c$
improves greatly the approximate eigenvalues. The fits with one
parameter and three parameters give also nearly the same results and
the behaviour of $\chi(\beta)$ for this case and the previous one
are very similar.
 
\begin{table}[htb]
\caption{\label{tab:fun2}Values of $\chi(\beta)$ for eigenvalues of
Hamiltonian~(\ref{eq:fungraph}) as a function of $\beta$, for various
parameterization of $N=b(\beta) n + l + c(\beta)$ (see table~\ref{tab:fun}).
$N^{(\textrm{C})}$, set 1 and set 3 give the exact result for
$\beta=\infty$.}
\begin{indented}
\item[]\begin{tabular}{@{}cccc}
\br
$\beta$ & $N^{(\textrm{C})}$ & Set 1 & Set 2 \\
\mr
0.5 & $0.17$ & $4.2\ 10^{-4}$ & $3.6\ 10^{-4}$ \\
1   & $0.15$ & $2.9\ 10^{-3}$ & $2.9\ 10^{-3}$ \\
2   & $0.11$ & $1.7\ 10^{-2}$ & $1.7\ 10^{-2}$ \\
\br
\end{tabular}
\end{indented}
\end{table}

\begin{table}[htb]
\caption{\label{tab:fun3}Eigenvalues $\epsilon(\beta_0,n,l)$ of Hamiltonian~(\ref{eq:fungraph}) with
$\beta_0=0.5$, for some sets $(n,l)$. 
First line: $\epsilon_{\textrm{num}}(\beta_0;n,\ell)$ from numerical integration; 
second line: $\epsilon^{(\textrm{f})}(\beta_0;n,l)$ given by (\ref{eq:epsfun}) with $N(\beta)$ defined by set 1;
third line: $\epsilon^{(\textrm{f})}(\beta_0;n,l)$ given by (\ref{eq:epsfun}) with $N(\beta)=N^{(\textrm{C})}$.}
\begin{indented}
\item[]\begin{tabular}{@{}ccccc}
\br
$l$ & $\epsilon(\beta_0;0,l)$ & $\epsilon(\beta_0;1,l)$ & $\epsilon(\beta_0;2,l)$ & $\epsilon(\beta_0;3,l)$ \\
\mr
0 & 0.39711 &   1.11714  &  1.64558  &  2.09628 \\
 & 0.42779 & 1.16223 & 1.68099 & 2.12205 \\
 & 0.26827 & 0.79105 & 1.15440 & 1.45987 \\
 \\
1 & 0.90598  &  1.45955  &  1.92580  &  2.34167 \\
 & 0.88794 & 1.46673 & 1.93564 & 2.34911 \\
 & 0.79105 & 1.15440 & 1.45987 & 1.73269 \\
 \\
2 & 1.25749  &  1.74247  &  2.17133  &  2.56288 \\
 & 1.23307 & 1.73892 & 2.17323 & 2.56506 \\
 & 1.1544 & 1.45987 & 1.73269 & 1.98358 \\
 \\
3 & 1.55457 &   1.99727  &  2.39917  &  2.77168 \\
 & 1.52908 & 1.98937 & 2.39764 & 2.77183 \\
 & 1.45987 & 1.73269 & 1.98358 & 2.21833 \\
\br
\end{tabular}
\end{indented}
\end{table}

\subsection{General considerations}
\label{subsec:gencons}

>From the results of this section, it is clear that a good choice for
the function $N(\beta)$ can greatly improve the accuracy of the energy
formulae. Unfortunately, the best functional form cannot be theoretically
predicted. In the cases studied in this paper, the same form can be given
to both coefficients $b(\beta)$ and $c(\beta)$ for a given Hamiltonian,
but with different parameters. The behaviour for $\beta=0$ and
$\beta\to\infty$ can sometimes be exactly computed. 

Obviously, the parameters for coefficients $b(\beta)$ and $c(\beta)$
depend on the points used for the fit but also on the particular choice
of functions (\ref{chi2}) and (\ref{chi2bis}). Other definitions--relative
error instead of absolute error, different chosen quantum numbers or
different chosen values of $\beta$ in summations--would have given other
numbers slightly different. It is worth noting that, for each case studied,
values of $\chi(\beta)$ obtained directly with coefficients 
$d_{\textrm{min}}(\beta)$
and $d_{\textrm{fit}}(\beta)$ for the set 2 are generally
very close. 

\section{Conclusions}
\label{sec:conclus}

The AFM was proposed in SSB as a tool to compute approximate 
analytical solutions of the Schr\"{o}dinger equation and then applied to 
the case of power-law radial potentials. The basic idea underlying this 
method is to replace an arbitrary potential $V(r)$, for which no 
analytical spectrum is known, by an expression of the type 
$\nu\, P(r)+g(\nu)$, $P(r)$ being a potential for which analytical 
eigenenergies can be found, $\nu$ the auxiliary field and $g(\nu)$ a 
well-defined function of this
extra parameter. This auxiliary field is such 
that its elimination as an operator leads to the original Hamiltonian. 
If $\nu$ is seen as a number however, analytical eigenenergies and 
eigenstates can be found, and the auxiliary field is eventually 
eliminated by a minimization on the eigenenergies. The approximation in 
the AFM comes thus from the use of the auxiliary field as a number 
rather than an operator.

In the present work, we have further investigated the AFM and obtained 
results that fall mainly in two categories: general properties of the 
AFM and analytical resolution of the Schr\"{o}dinger equation with 
potentials of the form $ar^\lambda\pm br^\eta$. Let us first summarize 
the general properties of the AFM that we have proved:
\begin{itemize}
\item The analytical expressions that are obtained for the 
eigenenergies by using the AFM preserve the general scaling laws 
(\ref{eq:scallawgenp}) in the case where $P(r)$ is homogeneous. Thanks to 
this feature, the number of relevant parameters in the Hamiltonian can be 
seriously reduced in order to simplify the problem.
\item Let $P_1(r)$ and $P_2(r)$ be two power-law potentials whose 
eigenenergies respectively read $E_1(N_1)$ and $E_2(N_2)$, 
where $N_1$ and $N_2$ are terms containing the radial and orbital 
quantum numbers. Then, if we apply the AFM to find the eigenenergies of 
potential $V(r)$ with both $P_1(r)$ and $P_2(r)$, the final results will 
have the same functional form, depending either on $N_1$ or on $N_2$ 
following the case. 
\item Perturbation theory can be reformulated within the AFM. If $\sigma$
is a small parameter and if $V(r)=\sigma v(r)$ 
is a potential that can be treated in perturbation, then 
$\left\langle v(r)\right\rangle$ can be equivalently replaced by 
$v(J(\nu_0))$ with $J(\nu_0)$ the average point defined by (\ref{eq:anu3}), 
at the first order in $\sigma$. Such a property could be useful in some cases 
since no integration is needed with the AFM.
\end{itemize}

Finally, it is worth summing up the results that we obtained by solving 
the Schr\"{o}dinger equation with potentials of the form 
$ar^\lambda\pm br^\eta$:
\begin{itemize}
\item Analytical formulae have been found for several potentials that 
are relevant in various domain of physics: anharmonic, quadratic plus Coulomb 
and funnel. To our knowledge, it is the first time that such formulae are found.
A comparison with the Kratzer and the quadratic plus 
centrifugal potentials which are analytically solvable is also performed.
\item By using the scaling laws, it appears that only one 
dimensionless parameter, denoted as $\beta$ in this work, `controls' the 
features of the various spectra. 
The approximate formulae for eigenenergies can be a complicated function of 
$\beta$ and of a number $N$ containing the radial and orbital quantum numbers.
This number is determined by the potential $P(r)$ chosen
and is \emph{a priori} independent of $\beta$. Nevertheless, 
a drastic improvement of the formulae 
we computed can be obtained by replacing this term $N$ by a function of the form 
$b(\beta)n+\ell+c(\beta)$, where $b(\beta)$ and $c(\beta)$ have to be 
fitted on the eigenenergies coming from a numerical resolution of the 
Schr\"{o}dinger equation. This has finally led us to very accurate 
formulae for all the various potentials we studied.
\end{itemize}

As an outlook, we mention that more complicated potentials could be 
studied with the AFM. In particular, we plan to apply this method to 
potentials involving an exponential, like the Yukawa potential for 
example. Such a work is in progress.  

\ack

CS and FB thank the F.R.S.-FNRS for financial support.

\end{document}